\newif\ifnotes
\notestrue 

\newif\ifaft
\aftfalse

\documentclass[a4paper,UKenglish,cleveref, autoref, thm-restate]{lipics-v2021}
\usepackage[table]{xcolor} 
\usepackage{makecell}

\usepackage[linesnumbered,ruled,vlined]{algorithm2e}
\DontPrintSemicolon
\newcommand{\Func}[2]{\text{#1(}#2\text{)}}
\usepackage{amsfonts}
\DeclareMathAlphabet{\mathbbmsl}{U}{bbm}{m}{sl} 
\usepackage{commath}

\usepackage[T1]{fontenc}
\usepackage{microtype}
\usepackage{pstricks}
\usepackage{graphicx}
\usepackage{amsmath}
\usepackage{amssymb}
\usepackage{xspace}
\usepackage{listings}
\usepackage{multirow} 
\usepackage{booktabs}
\usepackage{tabularx}
\usepackage{xspace}
\usepackage{multirow}
\usepackage{pifont}
\usepackage{amsthm}
\usepackage{wrapfig}

\usepackage[most]{tcolorbox}

\title{The Execution Dilemma in Pessimistic Blockchains: Profitability XOR Fair Ordering} 

\author{Artjom Pugatsov}{Delft University of Technology, the Netherlands}{apugatsov@gmail.com}{}{}

\author{Can Umut Ileri}{IOTA Foundation \and Delft University of Technology, the Netherlands}{canumutileri@gmail.com}{https://orcid.org/0000-0003-4136-9421}{}

\author{Jérémie Decouchant}{Delft University of Technology, the Netherlands}{j.decouchant@tudelft.nl}{https://orcid.org/0000-0001-9143-3984}{}

\authorrunning{A. Pugatsov, C. U. Ileri, J. Decouchant} 

\Copyright{Artjom Pugatsov, Can Umut Ileri, Jérémie Decouchant} 



\category{} 

\relatedversion{}
\relatedversiondetails{Full Version}{https://arxiv.org/abs/2604.23266}

\supplement{Source Code}
\supplementdetails[subcategory={Source Code}, cite={}, swhid={}]{Software}{https://github.com/Artjom-Pugatsov/genetic-sequencing}

\nolinenumbers 

\EventEditors{Aggelos Kiayias and Maria Kyropoulou}
\EventNoEds{2}
\EventLongTitle{8th Conference on Advances in Financial Technologies (AFT 2026)}
\EventShortTitle{AFT 2026}
\EventAcronym{AFT}
\EventYear{2026}
\EventDate{October 6--9, 2026}
\EventLocation{London, UK}
\EventLogo{}
\SeriesVolume{395}
\ArticleNo{10}

\begin{document}

\maketitle

\begin{abstract}
    The successive generations of consensus algorithms progressively displaced the performance bottleneck of blockchains to the execution layer. 
Recent works address the execution performance bottleneck by parallelizing the execution of non-conflicting transactions. 
Historically, execution closely followed the consensus-level transaction ordering determined by validators, a practice highly susceptible to Maximal Extractable Value (MEV) exploitation. Conversely, recent academic proposals introduce rigid fair-ordering protocols that are bound to severely restrict transaction reordering at the execution layer.
Parallel execution frameworks optimize the sequencing layer, which lies between consensus and execution
and assembles transaction batches from committed consensus blocks and transmits them to execution workers to maximize both execution parallelism and realized transaction fees. To achieve this optimization, current sequencing implementations may defer transactions. 
%
Importantly, these implementations do not maintain order-fairness properties. To the best of our knowledge, preserving these properties currently requires sequential execution, which would drastically reduce both profitability and performance. 

In this work, we address the tension between validator profit and order fairness using a dynamic optimization framework. We introduce a blockchain-agnostic model for transaction sequencing in a continuous setting where block sequencing and execution run concurrently. Consequently, when sequencing cannot be completed within the available time window, our framework dynamically returns its best intermediate result. Within this framework, we propose an anytime genetic algorithm that utilizes gas prices, object sets, and predicted execution times to optimize schedules. We also augment this algorithm to optionally maintain fair-ordering. 
We evaluate our approach with real-world datasets from Sui and Ethereum, and demonstrate that it increases validator profit by approximately 15\% and accelerates congestion relief speed by up to 58\%. Furthermore, we quantify the impact of fair-ordering constraints, showing that they can reduce validator profit by 50\% to 60\% during periods of high congestion. 
We provide the first evidence that enforcing strict fair ordering might effectively nullify the advantages of advanced sequencing.

\end{abstract}

\section{Introduction}

Historically, blockchain systems have faced significant performance limitations. For example, Bitcoin and Ethereum support only about 7 and 40 transactions per second (TPS), respectively~\cite{Li_Etherum_TPS, Corman_Bitcoin_TPS}. As blockchain applications have expanded beyond payments to domains such as decentralized finance and the Internet of Things, ensuring substantially higher throughput has become critical.
Early blockchains used monolithic architectures that tightly coupled transaction dissemination, consensus, verification, and execution, making it difficult to optimize individual components. This limitation led to lazy blockchain architectures~\cite{danezis2022narwhal,gao2022dumbo,spiegelman2022bullshark}, which separate these functions into distinct stages that can be optimized independently and executed concurrently. As a result, system performance is bounded by the slowest stage.
While research has mostly focused on consensus~\cite{Hao_Blockchain_Performance_analysis}, recent approaches, e.g., high-throughput Directed Acyclic Graph (DAG)-based protocols~\cite{keidar2021all_DAG_RIDER,Narwhal_and_tusk,babel2025mysticeti,polyanskii2025starfish}, have shifted the main performance bottleneck to the execution layer.

Execution performance can be optimized in two ways. Vertical optimization improves the execution of individual transactions, often by analyzing smart contract languages~\cite{garcia2017optimized, albert2022super}, but its effectiveness depends on transaction type. We focus on horizontal optimization, which increases throughput by executing independent transactions in parallel across multiple executors~\cite{pilotfish, ParBlockchain, Ruan_Execute-order-validate, cachin2016architecture_fabric, etherum_speedup, das2021efficientcrossshardtransactionexecution, Block-STM}. Using horizontal optimization, state-of-the-art execution engines reportedly achieve up to 170\,kTPS~\cite{Block-STM}.
The key property that enables efficient horizontal parallelization is transaction independence. Independent transactions can be executed concurrently across distributed workers, necessitating only a final state merge. In the object model~\cite{ezard2025nemo,SuiTouchedObjects, SolanaSpecifiedAccounts}, transactions consist of read and write operations on specific state objects and two transactions are therefore considered independent if their operations on common objects are read-only.

Traditional blockchains, such as Bitcoin and Ethereum, allow validators to choose transaction order within a block. This acts as a form of congestion control, as validators prioritize transactions with higher gas fees to mitigate spam and balance throughput with available execution resources~\cite{buterin2013ethereum}. However, this opaque discretion also enables Maximal Extractable Value (MEV), where validators reorder transactions for profit~\cite{daian2020flash}. In response, fair ordering protocols have been proposed for the dissemination and consensus layers~\cite{nasrulin2023accountable,kelkar2023themisbatch,zhang2020byzantinepompetimestamp}, enforcing transaction order based on arrival time. These approaches define two extremes: unrestricted validator control and rigid ordering that ignores validator incentives.

In this work, we investigate a middle ground based on the transaction sequencing framework of Sui~\cite{sui_congestion_control, sui_local_fee_markets}, which uses the object model. In this framework, transactions are either ordered for execution or selectively deferred to be reprocessed with later blocks when resource limits are exceeded. Sui orders transactions by gas price and processes them sequentially, deferring any transaction whose objects would exceed a protocol-defined execution capacity. 
We extend Sui's approach by incorporating execution-time estimates~\cite{predictin_gas_price,popovic2016estimating} and object-conflict information~\cite{static_analysis_eth_chahoki,tikhomirov2018smartcheck} directly into the sequencing process.  We specify a sequencing framework that supports the continuous execution model, where the processing of consecutive blocks by different blockchain layers might overlap, and changing congestion conditions. To address this specification, we propose an anytime genetic algorithm\footnote{An anytime algorithm returns a valid solution even if interrupted before completion.} that continuously refines the ordering as time permits.

As a summary, we make the following \textbf{contributions}:

$\bullet$ We introduce the first formalization of the transaction sequencing problem within a continuous execution framework. Unlike traditional single-shot approaches, our model explicitly accounts for overlapping block boundaries, transaction object conflicts, and dynamic scheduling deadlines under network congestion. We extend this model to formalize how order-fairness policies can be integrated as optional execution constraints.

$\bullet$ We design and implement an anytime genetic algorithm tailored for the sequencing layer to optimize validator profit using gas prices, object sets, and predicted execution times. The algorithm accommodates real-time execution constraints by immediately outputting a high-quality intermediate schedule if interrupted prematurely. We augment our genetic framework to support optional consensus-determined fair ordering. By strictly enforcing arrival-sequence constraints only for interdependent (conflicting) transactions, our algorithm intelligently allows the concurrent reordering of independent transactions.

$\bullet$ We implement our algorithm and compare it against several sequencing heuristics using real-world Sui data and synthetic Ethereum-based workloads. Our experiments quantify the impact of  perturbed execution-time estimates.
Using real-world blockchain data, our results demonstrate that the proposed genetic sequencer increases validator profit by up to 16\% under sustained congestion and accelerates congestion relief by up to 58\% during sporadic transaction spikes.

$\bullet$ We provide the first empirical evidence that enforcing strict fair ordering severely limits independent transaction optimization, imposing a 50\% to 60\% profit penalty on validators during peak congestion. Based on our model, algorithms, and datasets, our results suggest that rigid fair ordering may eliminate the performance benefits of advanced parallel sequencing layers.

\section{Background}

\begin{figure}[t]
    \centering
    \begin{subfigure}[c]{0.50\textwidth}
        \centering
        \includegraphics[width=\textwidth]{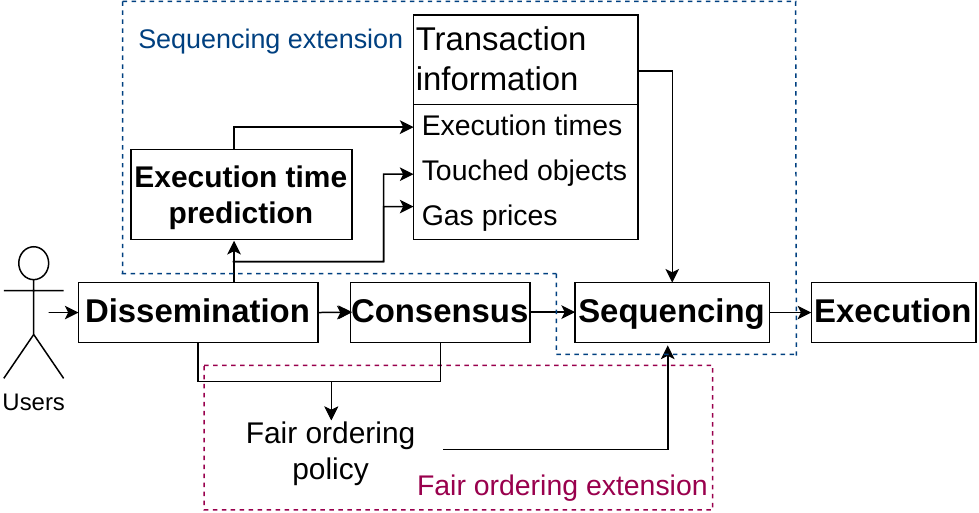}
    \end{subfigure}
    \hfill
    \begin{subfigure}[c]{0.43\textwidth}
        \centering
        \includegraphics[width=\textwidth]{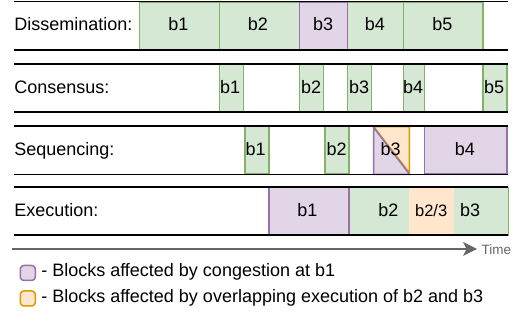}
    \end{subfigure}

    
    \begin{subfigure}[t]{0.48\textwidth}
        \caption{Lazy blockchain architecture}
        \label{fig:Lazy blockchain with sequencing}
    \end{subfigure}
    \hfill
    \begin{subfigure}[t]{0.48\textwidth}
        \caption{Continuous execution in a lazy blockchain}
        \label{fig:Parallel sequencing}
    \end{subfigure}
    
    \caption{Overview of the sequencing layer in a lazy blockchain. Fig.~\ref{fig:Lazy blockchain with sequencing} illustrates the modular architecture we consider, where sequencing bridges consensus and execution by computing schedules based on transaction gas prices, object sets, and predicted execution times, with optional fair ordering constraints. Fig.~\ref{fig:Parallel sequencing} shows the pipelined processing of blocks across blockchain layers, including two cases where the continuous execution requirement affects sequencing. 
    First, congestion detected during the execution of block 1 causes the dissemination of block 3 to end early, freeing time for additional sequencing of upcoming blocks 3 and 4. Second, an overlap between the execution of blocks 2 and 3 forces early termination of block 3's sequencing to maintain execution continuity.
    }
\end{figure}

    \subsection{Sequencing}


Fig.~\ref{fig:Lazy blockchain with sequencing} illustrates the modular architecture of a so-called lazy blockchain, where consensus and execution are decoupled. We focus on the sequencing layer, which receives blocks\footnote{We use the term ``block'' to denote a set of transactions sequenced and executed together. In DAG-based architectures, a single commit may include multiple vertices.} from the consensus layer. The sequencing layer assigns transactions to execution workers, manages congestion, and may defer transactions that cannot be completed within the available execution window.
We assume that execution-time estimates are available for all transactions. These estimates can be derived from transaction properties as early as during the dissemination phase, when transactions become known, although the estimation method itself is outside the scope of this work. We also consider an optional fair ordering constraint, which requires the sequencing layer to preserve the causal order of transactions.
The sequencing layer produces an estimated schedule, while the execution layer performs the actual execution. Because execution times may vary and conflicts can introduce additional delays, the realized execution order may differ from the estimated schedule.

\subsection{Scheduling Estimation}

While we focus on optimizing transaction ordering, evaluating a sequencing algorithm requires estimating the resulting execution schedule, which ultimately determines execution performance.
We assume a scheduling function that maps an ordered set of transactions to estimated start times and worker assignments. Although our method is independent of its specific implementation, we use a concrete scheduler to evaluate results.

Because optimal scheduling is NP-hard, we assume a deterministic polynomial-time greedy scheduler that processes transactions sequentially, assigning each to the earliest feasible position on a worker and deferring those that cannot be safely placed within a worker's schedule. Our approach extends Sui's to a fixed number of workers, whereas the original design assumed unbounded parallelism and focused solely on object availability. Alg.~\ref{pseudo_code:basic_multiworker_scheduler} details our extended scheduling algorithm.

\begin{algorithm}[t]
    \small
    \SetKwInOut{Input}{input}
    \SetKwInOut{Output}{output}
    \SetKwRepeat{Do}{do}{while}

    \Input{$L_{all}, d, m$}
    \Output{Assignment $A$, where $A[tx] = (t, w_{id})$ or $\text{None}$}
    \BlankLine
    $A \leftarrow \emptyset $ \\
    \For{$tx\: \textbf{in}\:L_{all}$}{
        $WrkFreeAt \leftarrow \emptyset $ \\
        \For(\tcp*[f]{find earliest free slot per worker}){$w\: \textbf{in}\: 0, 1, ..., m-1$} {
            $WrkFreeAt[w] \leftarrow  \Func{find\_earliest}{A, w, tx}$ \\
        }
        $wrk \leftarrow \Func{choose\_worker}{WrkFreeAt}$ \tcp*{choose worker based on policy} 
        \If{$wrk = \mathrm{None}$}{
           $A[tx] \leftarrow \text{None} $ \tcp*{no free workers} 
        }
        \Else{
           $A[tx] \leftarrow (WrkFreeAt[wrk], wrk) $ \tcp*{schedule at earliest available slot} 
        }
    }
    \Return $A$ 
    \caption{Greedy scheduler}
    \label{pseudo_code:basic_multiworker_scheduler}
\end{algorithm}

We use the Earliest Start Time policy, which assigns each transaction to the worker that allows the earliest execution and is the default policy in current blockchains. 
We further improve scheduling by allowing transactions to be inserted into gaps in existing schedules rather than only appended. While this improves utilization~\cite{ileri2025enhanced}, it increases complexity since gap structures must be maintained and queried.
For efficiency reasons, we store gaps in a B-tree ordered by duration and search from the smallest feasible gap upward, stopping at the first valid placement. This differs slightly from the original gap-filling formulation but achieves comparable schedule quality with better scalability for large workloads. We therefore adopt this greedy gap-filling scheduler as our evaluation scheduling function.

\subsection{Sequencing Objective and Requirements} 
\label{sec:sequencing-objectives}

\textbf{Objective.} We model sequencing as a combinatorial optimization problem that constructs a transaction schedule optimizing a given performance objective. 
For instance, minimizing makespan targets shortest completion time, while maximizing throughput focuses on the number of transactions executed within a fixed horizon. In this work, we aim at maximizing the total fees of scheduled transactions. 
A transaction's execution fee is calculated by multiplying its execution time by its gas price. This objective prioritizes less resource-intensive transactions while still admitting more congestive ones when their gas price compensates for their cost. It also aligns sequencing with validator incentives by maximizing expected revenue, encouraging continued participation in the network.

\textbf{Requirements.} The sequencing step should satisfy three requirements: high performance, adaptability, and continuous operation.
First, since sequencing is meant to improve execution performance, it should not delay it. 
%
Second, during low congestion periods, a simple sequencing strategy suffices and  spending excessive time on optimization would only increase latency. 
Differently, in periods of high congestion, a sequencer should dynamically increase its optimization effort and utilize an anytime algorithm to prioritize transaction parallelism, reordering independent transactions while deferring highly conflicting ones. 
Finally, sequencing must support continuity so execution can proceed to the next block without waiting for the previous one to fully complete. Instead, workers should continuously consume available transactions. As they finish, they immediately take on new work, potentially from later blocks, while earlier transactions may still be executing. This leads to fully overlapping execution across multiple blocks, with no strict barrier between them.

As illustrated in Fig.~\ref{fig:Parallel sequencing}, these requirements enable cross-layer parallelism. The system can dynamically adjust sequencing effort and timing under congestion, while still maintaining continuous execution and overlapping the processing of successive blocks. Sequencing can be stopped early to immediately feed execution, reducing idle time while allowing later optimization to refine future schedules.
We do not assume any specific coordination mechanism between layers. Instead, these requirements are reflected in the design of our ordering algorithm, which must be anytime, and inherently support continuous operation.

    \subsection{Greedy Ordering Baseline}

\begin{figure}[t]
    \centering
    \includegraphics[width=\textwidth]{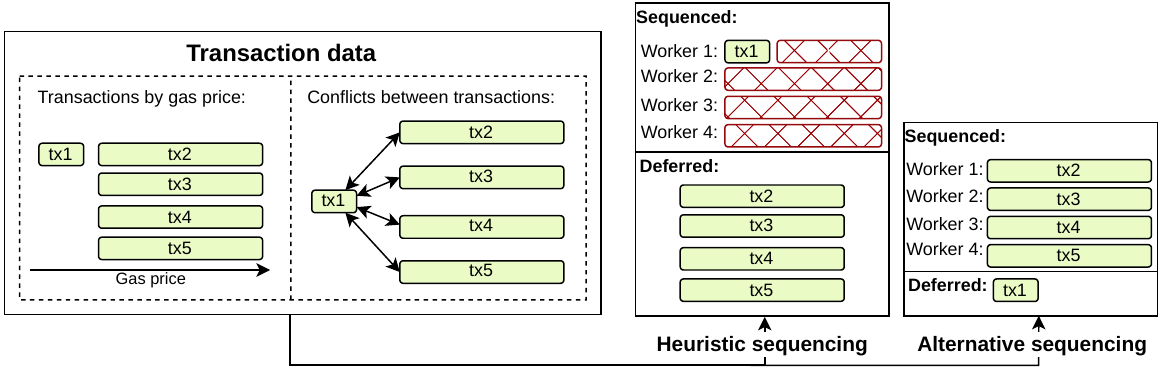}
    \caption{Greedy transaction sequencing based on gas price (left) compared to a feasible alternative (right). The width of each transaction corresponds to its execution time. The greedy algorithm is sub-optimal and defers transactions $tx2$--$5$, while the alternative algorithm only defers $tx1$.}
    \label{fig:Congestive transactions}
\end{figure}

Sui's current sequencing mechanism uses a heuristic that considers transactions in descending order of their gas prices.
Transactions are considered one at a time and either included or deferred based on their estimated impact on shared-object capacity.
This greedy approach can perform poorly under contention. A highly congestive transaction may be selected first, blocking many smaller transactions that could otherwise execute in parallel. As illustrated in the example of Fig.~\ref{fig:Congestive transactions}, scheduling transaction 1 first causes transactions 2 to 5 to be deferred, leaving execution workers underutilized. In contrast, deferring transaction 1 and scheduling transactions 2 to 5 increases parallelism, reduces deferrals, and may yield higher gas revenue.
Replacing gas price with another single ordering criterion does not resolve this issue, as a highly congestive transaction may still be scheduled first. Instead, we incorporate both execution-time estimates and transaction conflicts into the sequencing process to construct schedules that better utilize available parallelism and improve overall performance.

 \section{Modeling Continuous Transaction Sequencing}

In this section, we present what is, to our knowledge, the first formalization of sequencing within a continuous execution model. We begin by outlining our foundational assumptions, followed by the formalization of both single-shot and multi-shot continuous sequencing. Through these definitions, we explicitly establish the structural constraints that any valid sequencing algorithm, including ours, must satisfy.
We introduce our notation throughout the text as needed, and provide a comprehensive summary in Table~\ref{table:notation}. 

\begin{table}[t]
\centering
\caption{Notations.}
\label{table:notation}
\small
\begin{tabularx}{\columnwidth}{l X @{}}
\toprule
\textbf{Symbol} & \textbf{Description} \\
\midrule
\multicolumn{2}{@{}l}{\textit{Single-shot sequencing}} \\
\midrule
$d, m$ & Execution deadline ($d \in \mathbb{R}^+$), Number of workers ($m \in \mathbb{N}^+$) \\
$T_\text{all}, L_\text{all}$ & Set ($T_\text{all}$) and ordered list ($L_\text{all}$) of all transactions $\{tx_1, \dots, tx_n\}$ \\
${tx}_i$ & Transaction triplet $(O_i, t_i, g_i)$ with execution time $t_i$ and gas price $g_i$ \\
$O_i$ & Object accesses tuple $(R_i, W_i)$ where $R_i, W_i \subseteq \mathcal{I}$ are read and write sets \\
$\sigma$ & Scheduling function: $(L_\text{all}, d, m) \to \Pi_{L_\text{all}}$ \\
$\Pi_{L_\text{all}}$ & Schedule mapping: $tx_{i} \to \text{Either}\bigl(\texttt{None}, \, (s_i, w_i) \bigr)$ with start time $s_i$ and worker $w_i$ \\
$T_\text{sched}, P$ & Scheduled transactions subset ($T_\text{sched} \subseteq T_\text{all}$) and validator profit ($P \in \mathbb{R}^{+}$) \\

\midrule
\multicolumn{2}{@{}l}{\textit{Continuous sequencing (at block $i$)}} \\
\midrule
$d_i, d_{i_\text{new}}$ & Base execution deadline and extended deadline \\
$m_i, \theta$ & Number of workers and maximum deferral rounds ($\theta \in \mathbb{N}^+$) \\
$T_{i_\text{inc}}, T_{i_\text{all}}, T_{i_\text{sched}}$ & Incoming, total considered, and scheduled transactions subsets \\
$\delta_i, C_i^w$ & Maximum slack time and latest finish time of worker $w$ \\
$O_{tx_j}$ & Touched objects set $R_j \cup W_j$ for transaction $tx_j$ \\
$\alpha_k^{i}, S_i^{k}$ & Latest scheduled time for object $k$, and transactions subset touching object $k$ \\

\midrule
\multicolumn{2}{@{}l}{\textit{Fair ordering}} \\
\midrule
$L_\text{fair}, G_\text{fair}$ & Total transaction order list, and transaction dependency graph $(T_\text{all}, E)$ \\
\bottomrule
\end{tabularx}
\end{table}

    \subsection{Assumptions}

We consider a system of $m$ execution workers on which transactions are scheduled. 
We assume two conditions: (i) transaction read and write sets are available; and (ii) transaction execution times are known.

The first assumption is directly satisfied in the pessimistic execution model utilized by Sui and Solana, where each transaction must explicitly declare its read and write sets~\cite{SuiTouchedObjects, SolanaSpecifiedAccounts}. Even in environments without this structural requirement, static analysis tools can accurately predict or infer these sets beforehand~\cite{tikhomirov2018smartcheck, static_analysis_eth_chahoki}.
The second assumption is significantly more challenging to guarantee. Determining a transaction's exact execution path and runtime dynamically is notoriously difficult. Even with pre-execution, the actual processing time can vary based on its final position within a block. Nevertheless, extensive research into historical data analysis and predictive simulation has demonstrated high accuracy in estimating these runtimes~\cite{predictin_gas_price,popovic2016estimating}. In our experiments we explicitly introduce prediction inaccuracies to evaluate their impact on system performance.

\subsection{Single-Shot Sequencing}

\textbf{Input.} Let the execution deadline $d \in \mathbb{R}^+$ be the maximum amount of time a worker can be active for a given block. We consider a system that relies on $m \in \mathbb{N}^+$ workers to execute transactions in parallel. 
We are given a scheduling function $\sigma$ that takes as input an ordered list $L_{all} = [{tx}_1, {tx}_2, \dots, {tx}_n]$ of $n$ transactions and schedules all or part of them on the $m$ workers over a time period $d$, producing a schedule \(\Pi_{L_{all}}\).

A set of all transactions $T_{all} = \{{tx}_1, \cdots, {tx}_n\}$ consists of transactions ${tx}_i$ represented as a tuple $(O_i,t_i,g_i)$. The tuple \( O_i = (R_i, W_i) \) of object ids that ${tx}_i$ accesses consists of a set $R_i$ of object ids that it reads from and a set $W_i$ of object ids that it writes to. Variable $t_i \in \mathbb{R^{+}}$ is the  execution time of ${tx}_i$, and $g_i \in \mathbb{R^{+}}$ is the gas price of a unit of execution time paid by the transaction.

A scheduling function $\sigma:(L_{all}, d, m) \to \Pi_{L_{all}}$ is set in advance and represents a scheduler. It takes $L_{all}$, $d$ and $m$ as input and produces a schedule \(\Pi_{L_{all}}: tx_{i} \to \text{Either}\bigl(\texttt{None}, \, (s_i, w_i) \bigr)\). The tuple $(s_i, w_i)$ corresponds to the start time $s_i$ and the worker id $w_i$ on which the transaction has been scheduled. Transactions that are mapped to \texttt{None} are deferred to the following block.

\textbf{Constraints.} Any schedule \(\Pi_{L_{all}}\) produced by $\sigma$ must be subject to several constraints. First, a scheduled transaction ${tx}_i$ must be scheduled at a positive time (\( s_i \geq 0 \)) and adhere to the deadline (\(s_i + t_i \leq d \)). It also must be assigned to a valid worker id \(w_i \in \{0, 1, \cdots, m-1\}\). Finally, the execution must be conflict-free: if ${tx}_j$ is scheduled on the same worker as ${tx}_i$ $(w_i = w_j)$ or if the transactions conflict i.e., 
\(
((W_i \cap (R_j \cup W_j)) \neq \emptyset) \lor ((W_j \cap (R_i \cup W_i)) \neq \emptyset)
\) then their executions must not overlap (\(s_j + t_j \leq s_i \:\lor \:  s_i + t_i \leq s_j\)). 

\textbf{Objective.}
Our goal is to find an ordering $L_{all}$ of transactions $T_{all}$, such that for a given $\sigma$ we maximize the validator profit \(
P := \sum_{(O_i,t_i,g_i)\in T_{sched}}t_i\cdot g_i
\)
where \(T_{sched} := \{ tx_i \in T_{\text{all}} \mid \Pi_{L_{all}}(tx_i) \neq \text{None} \}\)  is the set of scheduled transactions.

\subsection{Multi-Shot Continuous Sequencing}

Our single-shot sequencing formulation captures the transaction sequencing one block at a time. But as opposed to consensus, which operates in discrete intervals, transaction execution is a continuous process: there is no need to wait for all the transactions of a previous block to be executed before executing non-conflicting transactions from the next block. Our goal in this section is to turn this isolated, single-block problem into one whose parameters are shaped by the outcome of the previous block: the only new input carried over between iterations is the initial busy status of the workers, while every other effect of continuity is captured through changes to the problem's parameters. To allow for transaction execution continuity, we extend our formulation of sequencing in several ways.

We label parameters based on the block number for which we generate the sequence: $T_{i_\text{inc}}$, $d_i$, $m_i$ are respectively the set of transactions, execution deadline and the number of workers at block $i$. $T_{i_\text{inc}}$ denotes the set of new transactions that just came in during the new round. The full set of transactions that will be scheduled in round $i$ is denoted $T_{i_\text{all}}$. We also define $\theta$, a parameter which determines the maximum number of rounds for which a transaction can be deferred before it is canceled.

We cancel transactions that have been deferred for too many rounds, and add the rest to the transaction set of the block 
    \(T_{i_\text{all}} = T_{i_\text{inc}} \:\cup\: \{tx_j \mid \Pi_{L_{i-1}}(tx_j) = \text{None} \:\land\:  tx_j \notin T_{(i-\theta)_\text{inc}} \}\).

We extend the deadline by the difference between the time the earliest worker is free and the deadline at the previous block: let $\delta_i$ be the maximum slack time at the end of round $i$, 
\(\delta_i = \max_{w \in [0,1,..m_i-1]} \left( d_{i_\text{new}} - C_i^w \right)\),
where $ C_i^w$ is the latest execution end time for worker $w$ at round $i$
    \(C_i^w = \max \{ s_j + t_j \mid \Pi_{L_i}(tx_j) = (s_j, w)\: , \: {tx_j \in T_{i_\text{sched}}}\} \).
We define the extended deadline to be \( d_{i_\text{new}} = d_i + \min(\delta_{i-1}, d_{i-1}) \).
We limit the maximum additional execution time per round by the execution time of the previous round, to disallow propagation of multiple rounds.

To ensure that transactions cannot be scheduled before the executor is fully done with transactions from a previous block we introduce a constraint:  if a transaction $tx_j$ from block $i$ is scheduled at \( \Pi_{L_{i}}(tx_j)  = (s_j, w_j) \), then 
    \(s_j \geq \delta_{i-1} -(d_{i-1} - C_{i-1}^{w_j})\)

Finally, we have to make sure that transactions do not access objects before all the transactions that read from or write to the overlapping objects of the previous block have finished: let $O_{tch_j} = R_j \:\cup\:W_j$ for a given $tx_j\in T_i$, then $\alpha_k^{i-1}$ is the latest time at which a transaction touching object $k$ has been scheduled in round $i-1$:
    \( \alpha_k^{i-1} = \max_{tx_j \in S_i^k} \bigl( s_j + t_j \bigr) \)
with $S_i^k$ the set of transactions that touch the object $k$ in round $i$:
    \(S_i^{k} = \{ tx_j \in T_{i_{sched}} \mid k \in O_{tch_j} \}\).
Then a transaction can only be scheduled after the latest time that an object has been freed in the previous round, so if a transaction $tx_j$ from block $i$ is scheduled at \( \Pi_{L_{i}}(tx_j)  = (s_j, w_j) \) , then
\(s_j \geq \delta_{i-1} - (d_{i-1} - \max_{k \in O_{tch_j}}(\alpha_k^{i-1}))\).

This approach globally aligns the start times for the new block with the worker that is free the earliest in the previous schedule, while shifting other executors by how late their execution is compared to the earliest executor. 

The main benefit of this approach is that it enables the continuity of execution without major changes to the problem structure. The model remains the same, barring the required addition of new hard-coded transactions that represent objects and workers not being free at the start of the round. This allows us to handle sequencing of each block as an independent combinatorial problem and additionally enables varying system parameters, such as $d$, $m$ or $\sigma$, dynamically based on the demand.

\subsection{Fair Ordering}

So far, our formalization of sequencing can arbitrarily reorder transactions. A strictly opposite approach would enforce fair ordering~\cite{kelkar2023themisbatch,zhang2020byzantinepompetimestamp,segalini2026tilikum,putnik2026herring} and focus on maintaining an order of transactions closely aligned with the order in which they have been submitted to the mempool. We extend our continuous sequencing model to include this constraint in order to measure its impact on the performance.

To do so, we modify our models in two ways. 
First, $T_\text{all}$ is replaced by $L_\text{fair} = [{tx}_1, {tx}_2, \cdots, {tx}_n] $ which corresponds to the total order over transactions produced by fair ordering.
Additionally, the scheduling function must respect the causal order established by this order within $L_\text{fair}$, meaning that for every two conflicting transactions ${tx}_i, {tx}_j$ where $i < j$: if the transaction positioned later in the order is scheduled, i.e., $\Pi(tx_j) = (s_j, w_j)$, then the earlier transaction must also be scheduled before it, i.e., $\Pi(tx_i) = (s_i, w_i)$ and $s_i + t_i \leq s_j$.

The scheduling algorithm must also be adapted to satisfy the fair ordering constraint. To preserve a simple and greedy design, we assume that the transactions passed to the scheduler are first totally ordered in a way that respects the causal order derived from $L_\text{all}$. As a result, the set of candidate schedules available to the scheduler is restricted. Specifically, this set consists of all possible traversals of the dependency graph $G_\text{fair}$ constructed from $L_\text{fair}$.

More specifically, the scheduling algorithm is modified in the following ways. First, if a transaction is deferred, all conflicting transactions that are ordered after the deferred one are also deferred. Second, each transaction can only be scheduled after all conflicting transactions that were ordered before it.

The pseudocode is shown in Algorithm~\ref{pseudo_code:fair_multiworker_scheduler}. Note that the inner workings of the find\_earliest function have to be changed to return a time greater than the end time of any previously scheduled conflicting transaction. 

\begin{algorithm}[t]
    \small
    \SetKwInOut{Input}{input}
    \SetKwInOut{Output}{output}
    \SetKwRepeat{Do}{do}{while}

    \Input{$L_\text{all}, d, m$}
    \Output{Assignment $A$, where $A[tx] = (t, w_{id})$ or $\text{None}$}
    \BlankLine
    $Def \leftarrow \emptyset $ \tcp*{deferred txs that will be added to the next block} 
    $A \leftarrow \emptyset $ \\
    \For{$tx\: \textbf{in}\:L_\text{all}$}{
        \If(\tcp*[f]{$tx$ conflicts with prev.\ deferred}){$\mathrm{does\_conflict\_with\_any}(tx, Def)$}{
           $A[tx] \leftarrow \text{None} $ \\
           $Def \leftarrow Def \cup \{tx\} $ \tcp*{defer tx}
           \textbf{continue}\\
        }  
        $WrkFreeAt \leftarrow \emptyset $ \\
        \For(\tcp*[f]{find earliest free slot per worker}){$w\: \textbf{in}\: 0, 1, \cdots, m-1$} {
            $WrkFreeAt[w] \leftarrow \Func{find\_earliest}{A, w, tx}$ 
        }
        $wrk \leftarrow \Func{choose\_worker}{WrkFreeAt}$ \tcp*{choose worker based on policy}
        \If(\tcp*[f]{no free worker}){$wrk = \mathrm{None}$}{
           $A[tx] \leftarrow \mathrm{None} $ \\ 
           $Def \leftarrow Def \cup \{tx\} $ 
            \tcp*{defer tx}
        }
        \Else{
           $A[tx] \leftarrow (WrkFreeAt[wrk], wrk)$ \tcp*{schedule at earliest available slot} 

        }
    }
    \Return $A$ 
    \caption{Order-fair greedy scheduler}
    \label{pseudo_code:fair_multiworker_scheduler}
\end{algorithm}

\section{Genetic Sequencing Algorithm}

Rather than optimizing for an ideal scheduling function, a problem that could be formulated for a constraint solver, we opt against this approach due to its  computational inefficiency. As noted by Chahoki et al.~\cite{chahoki2025conthereum}, using a constraint solver would introduce a new performance bottleneck, ultimately diminishing the throughput benefits of sequencing.

Instead, we focus on transaction ordering under a deterministic scheduling algorithm. This design choice explicitly accounts for imperfect information: the sequencer only has access to execution time estimates, which inherently slightly deviate from actual runtimes in practice, resulting in discrepancies between predicted and actual schedules~\cite{predictin_gas_price}.

Under this deterministic model, the most straightforward approach is to utilize a heuristic to order transactions. Because this problem closely resembles a multidimensional extension of the knapsack problem~\cite{sequencing_as_knapsack}, sorting transactions by gas price serves as our baseline heuristic. For comparison, we also evaluate three alternative heuristics: (i) a randomized order; (ii) the historical arrival order (the sequence in which transactions appeared in the source data); and (iii) shortest execution time first.

\subsection{Genetic Ordering}

\begin{wrapfigure}{r}{8cm}
    \includegraphics[width=8cm]{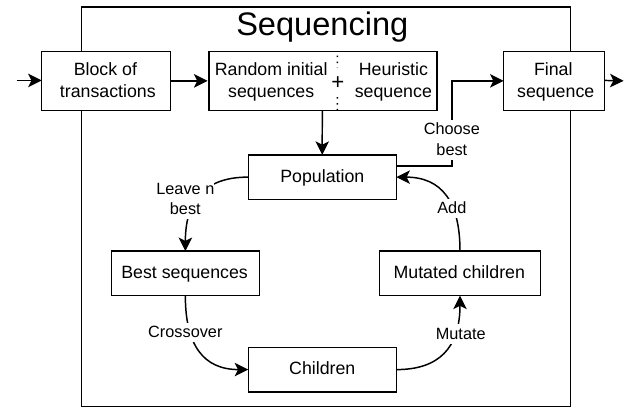}
    \caption{High level overview of our genetic transaction sequencing algorithm.}
    \label{fig:Inner workings of genetic sequencingg}
\end{wrapfigure} 

By fixing the scheduling function, the sequencing problem maps naturally onto a genetic algorithm framework. Here, each candidate solution is encoded as a permutation of the available transactions, with its sequence mapping directly to the scheduler's processing order. The fitness function is determined by applying the scheduling function and calculating the resulting profit. As illustrated in Fig.~\ref{fig:Inner workings of genetic sequencingg}, the algorithm initializes a population of transaction sequences and iteratively optimizes them through crossover and mutation operations, retaining only the fittest half of the population at each generation.

Because the gas price heuristic often yields high-quality baselines in practice, we seed our initial population with this heuristic ordering rather than starting from a random ordering. Consequently, our approach shifts the genetic algorithm's objective from an expensive global optimum search to a highly localized refinement of an already strong initial solution.




 Alg.~\ref{pseudo_code:genetic_alg_core} details the pseudocode of our genetic sequencing algorithm, which uses seeding, elitism, order crossover (OX) and insertion mutation, which have been shown to be effective for scheduling problems~\cite{CHENG1996983_genetic_for_sched_best_practices}. We discuss these techniques in the following:

\begin{algorithm}[t]
    \small
    \SetKwInOut{Input}{input}
    \SetKwInOut{Output}{output}
    \SetKwRepeat{Do}{do}{while}

    \Input{$budget, pop\_size, T_\text{all},  \sigma, d, m $}
    \Output{$L_\text{all}$}
    \BlankLine
    $L_\text{init} \leftarrow \Func{highest\_gas\_price\_order}{T_\text{all}}$ \tcp*{seeding}
    $P_\text{init}\leftarrow \Func{random\_sequences}{T_\text{all}, pop\_size} \cup \{L_\text{init}\}$\\
    $P_\text{eval} \leftarrow \Func{eval}{P_\text{init}, \sigma, d, m}$ \\
    $budget \leftarrow budget - (pop\_size + 1)$\\
    \While{$budget > 0$}{
        $P_\text{kids}\leftarrow \Func{generate\_kids}{P_\text{eval}, pop\_size}$ \tcp*{OX}
        $P_\text{kids}\leftarrow \Func{mutate}{P_\text{kids}}$ \tcp*{insertion}
        \For{$kid \in P_\text{kids}$}{
          $P_\text{eval} \leftarrow P_\text{eval} \cup  \{\Func{eval}{kid, \sigma, d, m}\}$\\
          $budget \leftarrow budget - 1$\\
        }
        $P_\text{eval} \leftarrow \Func{order\_by\_profit}{P_\text{eval}}$\\
        $P_\text{eval} \leftarrow \Func{limit\_population}{P_\text{eval}, pop\_size}$\tcp*{elitism}
    
    }
    \Return $P_\text{eval}[0]$ 
    \caption{Basis of our genetic sequencing algorithm}
    \label{pseudo_code:genetic_alg_core}
\end{algorithm}

\begin{itemize}
    \item \textbf{Seeding.} An initial solution is added to the starting population. This solution is produced heuristically by ordering transactions based on their gas price~\cite{seeding}.
    
    \item \textbf{OX.} To generate a child from two parent solution sequences, a random slice (random location and random length) is taken from one sequence and is inserted into another sequence at the same spot. Then the resulting sequence is scanned, and the possible second occurence of a transaction is removed~\cite{OX}.
    
    \item \textbf{Insertion mutation.} To mutate a sequence, a random transaction is removed and inserted into a different position~\cite{Insertion_mutation}.
    
    \item \textbf{Elitism.} A $(\mu + \lambda)$ survival strategy is employed. This means that each iteration parents $(\mu)$ and children $(\lambda)$ are combined and among them a new population is formed by choosing the sequences with the highest fitness~\cite{elitism}.
\end{itemize}

Seeding and elitism help to ensure faster convergence. To slightly offset this, we found it beneficial to set the mutation probability to $60\%$. The result is that the genetic algorithm acts similarly to a local search procedure, with a higher degree of exploration within the earlier iterations. Initially, an explicit local search procedure was also tested. It provided little benefit to the solution quality compared to extending the budget for the genetic algorithm, while requiring a lot of sequence evaluations.
In our experiments, we use $pop\_size=100$, $budget\in\{1000,5000\}$, $m=4$. The value we use for parameter $d$ depends on the dataset ($62,500$ for Sui and $1,250,000$ for ETH).  

\subsection{Adding Fair Ordering}

When adapting the genetic algorithm to maintain a fair order established by the consensus layer, we have to ensure that the transaction encoding represents a valid total order that respects the causal relations specified by $L_\text{fair}$. 
To do so, we apply the following changes. First, we change the initial seed $L_\text{init}$ to correspond to the sequence obtained through a topological traversal of $G_\text{fair}$ by prioritizing transactions with a higher gas price.
Then, we generate the rest of the initial parents $P_\text{init}$ by traversing $G_\text{fair}$ uniformly at random.
We modify the crossover operation so that it iterates over the parents and selects the first transaction whose dependencies are satisfied. The process continues until a valid total order consistent with $G_\text{fair}$ is generated.
We also modify the mutation operator as follows. First, for a randomly selected transaction, we determine its maximal neighborhood of non-conflicting transactions within the current total order by identifying the closest conflicting transactions before and after it. Then, we relocate the selected transaction to a position that is chosen uniformly at random within this neighborhood.

\section{Performance Evaluation}


\subsection{Settings}

In our evaluation, we do not report the runtime of the algorithms, and instead focus on execution performance. The scheduling algorithms are allowed to run until the next block is generated and interrupted if needed. The genetic algorithm is an anytime algorithm, so result quality depends on the allotted time. We do not estimate the possible real-world execution time. Instead, we report it in terms of iterations of the estimated sequencer. For reference, on the continuous Sui workload, median per-round wall-clock time stayed under $300$\,ms for all heuristic sequencers, versus roughly $1.0$\,s for GE10 and $4.4$\,s for GE50. The runtime of the genetic sequencer is proportional to the number of total evaluations of the estimated sequencer. The number of evaluations is equal to the number of children per epoch multiplied by the number of epochs. Additionally, the evaluations of children within a single epoch can be done in parallel. To measure the impact of the allotted execution time on performance, we do two runs of the genetic sequencer with the population of $100$: for $10$ and $50$ epochs, corresponding to $1,000$ and $5,000$ iterations of the heuristic sequencer, respectively. This allows us to quantify the benefit of additional computation time.

We focus on two scenarios with different solution quality metrics: sustained congestion and a single spike in congestion. In the sustained congestion scenario, we have congestion for the whole duration of sequencing and measure solution quality as the sum of gas prices paid by the scheduled transactions. In the single spike scenario we have congestion for a period of ten blocks and determine the solution quality by how quickly the sequencers managed to clear congestion. We define congestion clearance as the index of the first block after the spike period that contains no deferred transactions.

\subsection{Datasets and Parameters}

The performance of our sequencing approach is highly dependent on the structure and parameters of the transactions to be scheduled. To obtain measurements that are representative of the expected real-world performance, we base our datasets on two widely used blockchain systems that support smart contracts: Sui and Ethereum. We took transactions that were executed within a single day on both chains and streamed them to different sequencers. For Ethereum, we took the transactions from block 23359822 up to block 23369821, and for Sui we used transactions from epoch 840. Both represent recent active network conditions and span a sufficient transaction volume to capture realistic congestion dynamics.


To evaluate sequencing algorithms, we extract or generate the following parameters for a transaction ${tx}_i$:
\begin{itemize}
    \item Touched objects ($O_i$) - object set of transactions.
    \item Execution time ($t_i$) - amount of computation the transaction takes.
    \item Gas price ($g_i$) - price that the party submitting the transaction is paying per execution unit of the transaction.
\end{itemize}

The following parameters are independent of transaction data and control how congested the blockchain is. We vary them across experiments to simulate different scenarios.
\begin{itemize}
    \item Transactions per block ($|T_{inc}|$) - the number of transactions submitted for one block.
    \item Maximum execution time ($d$) - the maximum execution time that a single worker can work for per one block.
    \item Maximum number of deferrals ($\theta$) - the maximum number of blocks for which a transaction can be deferred before it gets canceled.
    \item Number of executors ($m$) - the number of executors that can execute the transactions in parallel.
\end{itemize}

All transactions are initially ordered according to their original positions within the blockchains to better capture temporal fluctuations in gas price trends, demand, and potential conflicts between contemporaneous transactions that may access overlapping sets of objects. To construct a continuous sequence of blocks, we partition the ordered transaction sequence into consecutive groups according to a predefined block size.

We map platform-specific parameters to the general model as summarized in Table~\ref{table:Parameters used}. Execution time was derived from transaction effects for Sui and transaction receipt for Ethereum. The gas price was taken from the transaction data for Sui and from the receipt for Ethereum. Touched objects were determined from changed and unchanged objects specified in transaction effects for Sui and were not collected for Ethereum due to them not being explicitly listed in the available data and requiring high-overhead trace analysis of transactions. We explain below how we add object sets to Ethereum transactions.

\begin{table}[t]
\centering
\caption{Derivation of model parameters from real-world data.}
\label{table:Parameters used}
\smaller 
\begin{tabular}{@{}lll@{}}
\toprule
Parameter      & Sui Dataset                & Ethereum Dataset            \\ \midrule
Execution Time & Transaction Effects       & Transaction Receipt        \\
Gas Price      & Transaction Data          & Transaction Receipt        \\
Touched Objects & Shared Objects & N/A (Account-based)        \\ \bottomrule
\end{tabular}
\end{table}

We note that Ethereum uses a base fee system to determine the minimum gas price. Therefore, to better compare transactions across different time periods, we normalize the reported gas used value by this base fee. To do this, we divide the gas price by the base fee for a given block made during different time periods.

\begin{figure*}[t]
    \centering
    \includegraphics[width=0.9\textwidth]{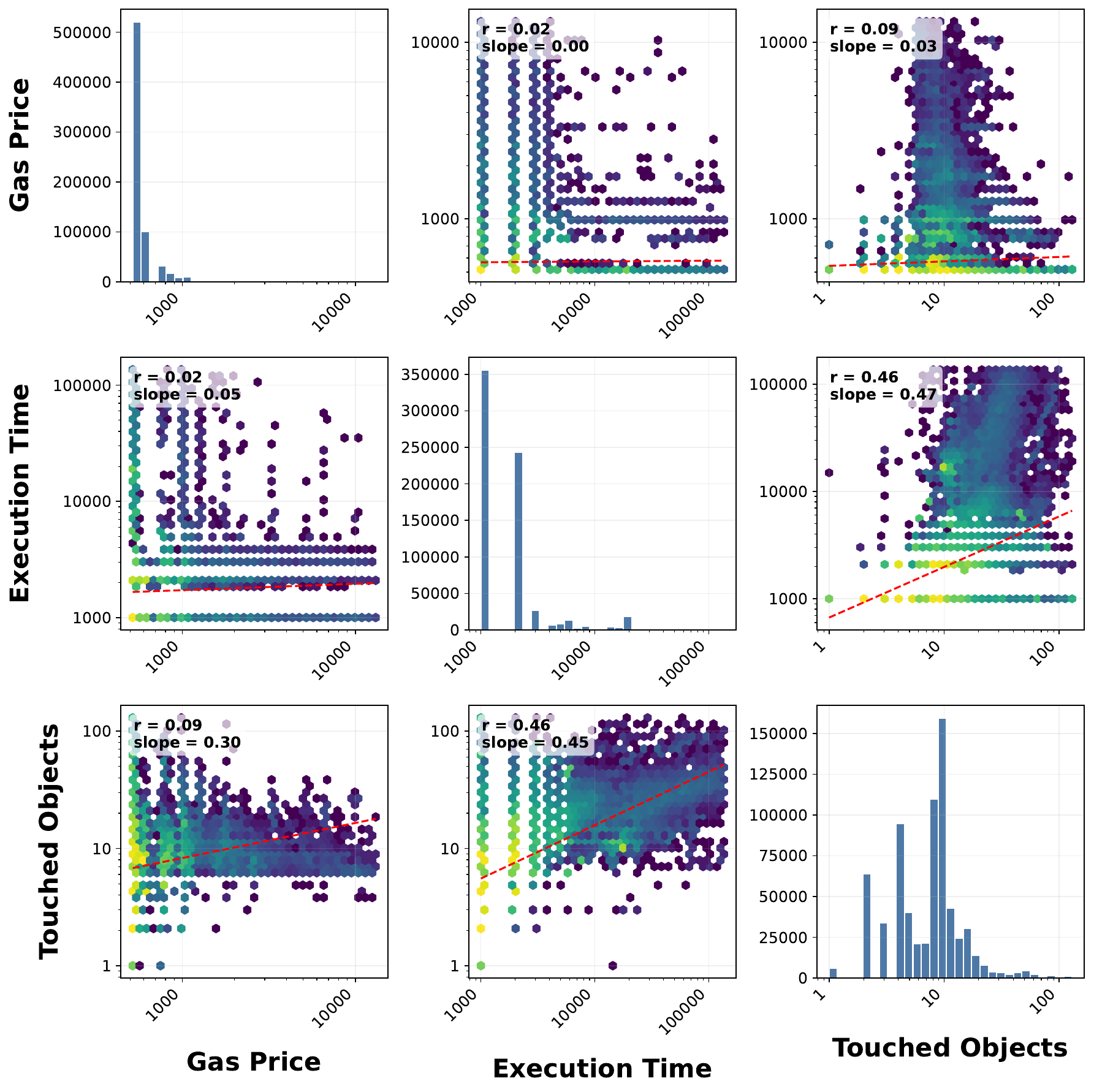}
    \caption{Log-scaled distribution and density plots for gas price, execution time, and total number of touched objects per transaction for SUI data. The units of these metrics are defined by the specific blockchain. Plots include a least-squares regression fit (slope is indicated) and Pearson correlation (r) for each pair of variables. Plots on the diagonal report the counts for each metric, other plots report the value of a metric depending on another one (e.g., number of touched objects depending on gas price). }
    \label{SUI data correlation}
\end{figure*}

\begin{figure*}[t]
    \centering
    \includegraphics[width=0.65\textwidth]{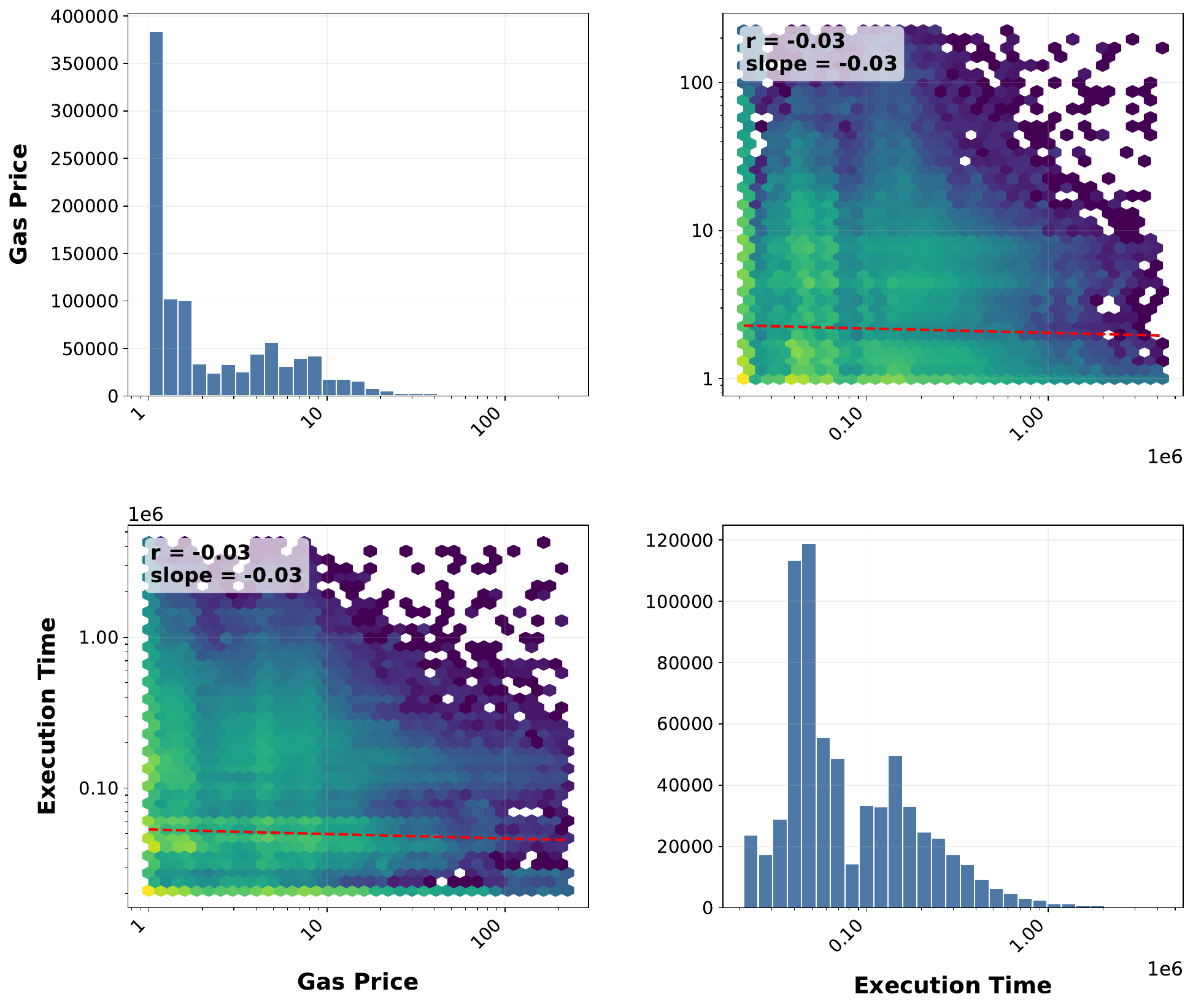}
    \caption{Log-scaled distribution and density plots for gas price and execution time for Ethereum data. Plots include a least-squares regression fit (slope is indicated) and Pearson correlation (r) for each pair of variables.}
    \label{ETH data correlation}
\end{figure*}

Since the results heavily depend on the data used, we want to first demonstrate the structure of the data and examine the distribution of values, as well as show their correlation with each other. Fig.~\ref{SUI data correlation} demonstrates the distribution of values for the Sui dataset, and Fig.~\ref{ETH data correlation} shows the distribution of values for the ETH dataset. From the datasets, we can see that the only two significantly correlated parameters are the number of touched objects and the execution time of transactions ($r=0.45$), while neither execution time nor number of touched objects is significantly correlated with gas price.  

As we mentioned before, we could not infer the accounts that were touched by  Ethereum transactions. Instead we resampled Sui's dataset. To construct the Ethereum-based dataset, we projected Sui's data on object conflicts over Ethereum's execution and gas price records. Specifically, we picked a random point in the Sui dataset and appended this information to the Ethereum stream one transaction at a time, preserving the order in which both datasets were scheduled on the blockchain. This approach introduces a key limitation: it removes the correlation between the number of touched objects and the execution time. For this reason, our Ethereum's dataset is partially synthetic and we refer to it as the ``Ethereum-based" dataset.

\begin{wrapfigure}{l}{8cm}
    \includegraphics[width=0.45\textwidth]{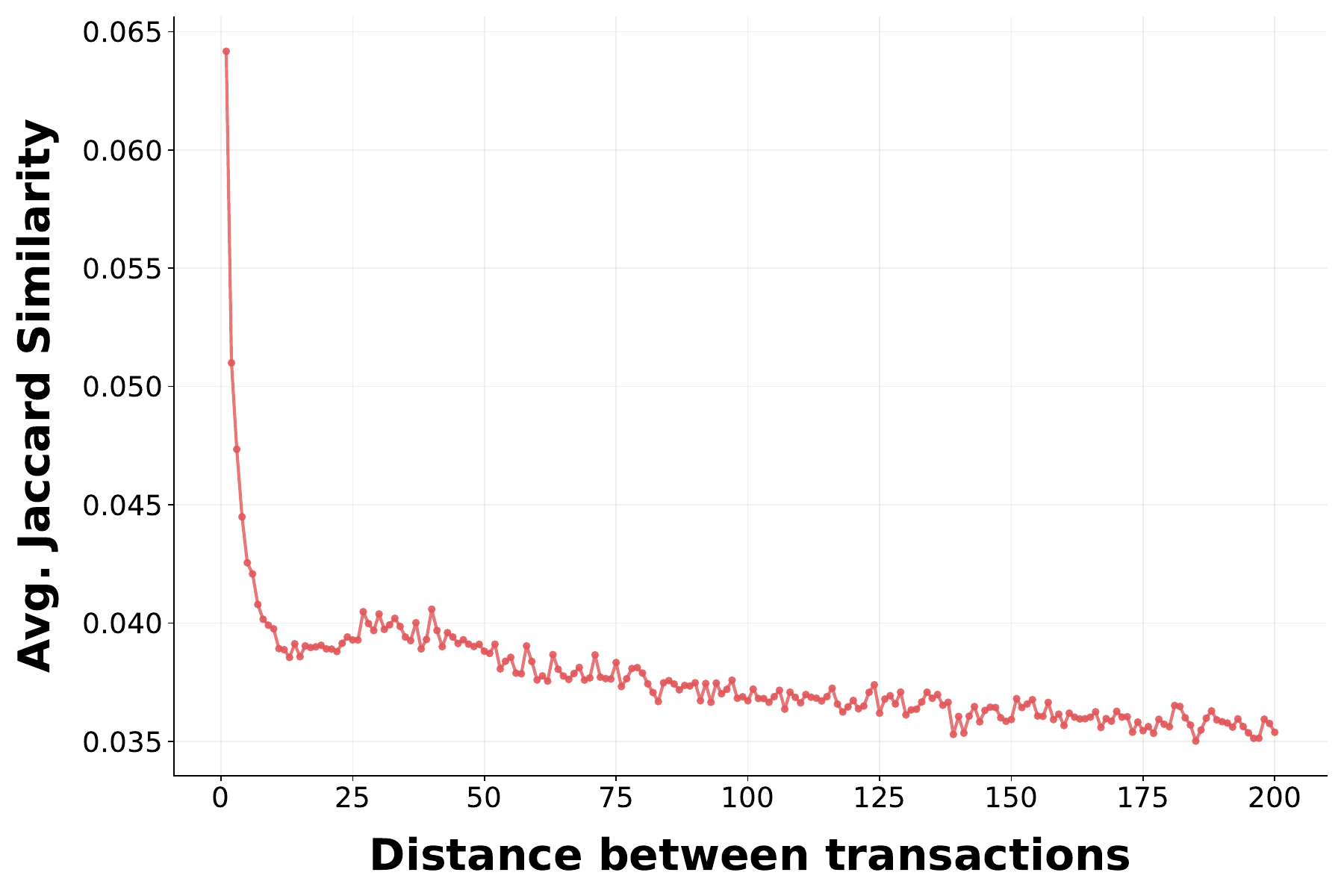}
    \caption{
    Average Jaccard similarity between transaction object sets as a function of execution-time distance, computed from a random sample of 50,000 transactions.}
    \label{SUI tx overlap}
\end{wrapfigure}

As shown in Fig.~\ref{SUI tx overlap}, we measured to what degree transactions overlap depending on the order in which they were submitted. The average Jaccard similarity follows a power-law decay as a function of transaction distance, with a sharp initial drop and a heavy tail. This confirms the significance of temporal locality and also shows the presence of several frequently touched global objects. 

The prevalence of temporal locality implies that a realistic dataset must consist of locally dependent transactions. It cannot be represented by independent sampling which would lead to a much lower degree of transaction conflicts. Furthermore, the heavy tail indicates the existence of heavily congested shared objects and reaffirms the need for congestion control.

\subsection{Sequencing Baselines}

We evaluated five different sequencing algorithms that rely on heuristics. All of them, except for the one labeled as "Sui", additionally use the gap-filling modification. The sequencers are as follows:

\begin{itemize}
    \item Sui - sequencer that orders transactions by descending gas price and does not use gap-filling.
    \item Gas Price (GP) - sequencer that orders transactions by descending gas price and does use gap-filling.
    \item Lowest Execution Time (LET) - sequencer that orders transactions by ascending execution time.
    \item Given - sequencer that orders transactions according to the order in which they have been executed on the blockchain.
    \item Random - sequencer that shuffles the order of transactions.
\end{itemize}

We use two versions of our genetic sequencing algorithm with two sets of parameters to demonstrate the impact of search depth on solution efficiency.
\begin{itemize}
    \item Genetic 100/50 (GE50) - genetic sequencer with population of 100 that runs for 50 epochs.
    \item Genetic 100/10 (GE10) - genetic sequencer with population of 100 that runs for 10 epochs.
\end{itemize}

\subsection{Sustained Congestion}


Unless stated otherwise, we use $m=4$ executors, a maximum of $5$ deferrals, $150$ transactions per block, and an execution deadline of $62{,}500$ time units, yielding an inclusion rate of approximately $50$--$60\%$ for the GP sequencer. Simulations run for $200$ blocks, and we report the cumulative validator profit normalized to the worst-performing sequencer. To highlight steady-state behavior, results are shown from block $20$ onward.

\begin{figure}[t]
    \centering
    \begin{subfigure}[b]{0.45\textwidth}
        \centering
        \includegraphics[width=\textwidth]{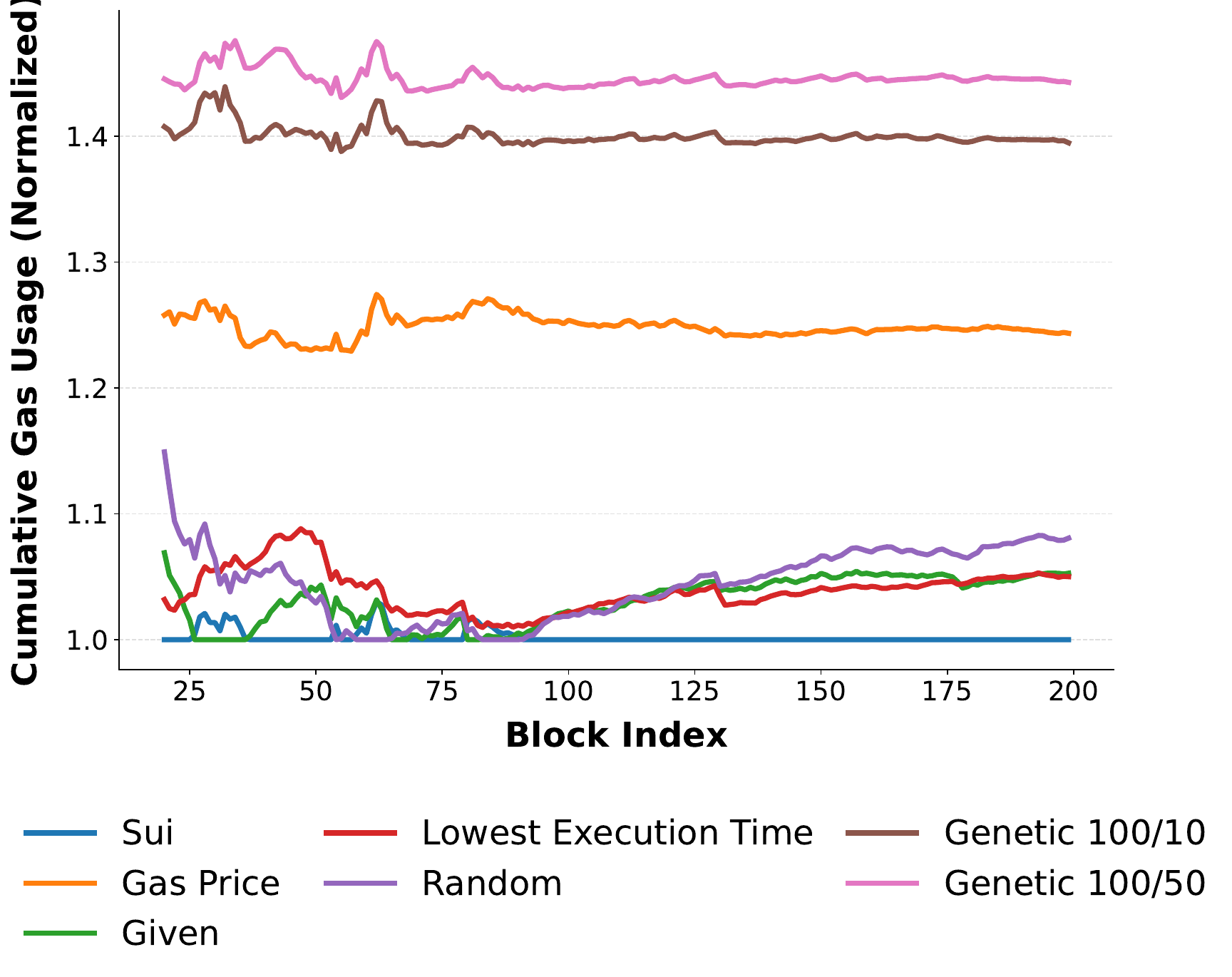}
        \caption{Sui dataset}
        \label{Validator profit under heavy load Sui}
    \end{subfigure}
    \hfill
    \begin{subfigure}[b]{0.45\textwidth}
        \centering
        \includegraphics[width=\textwidth]{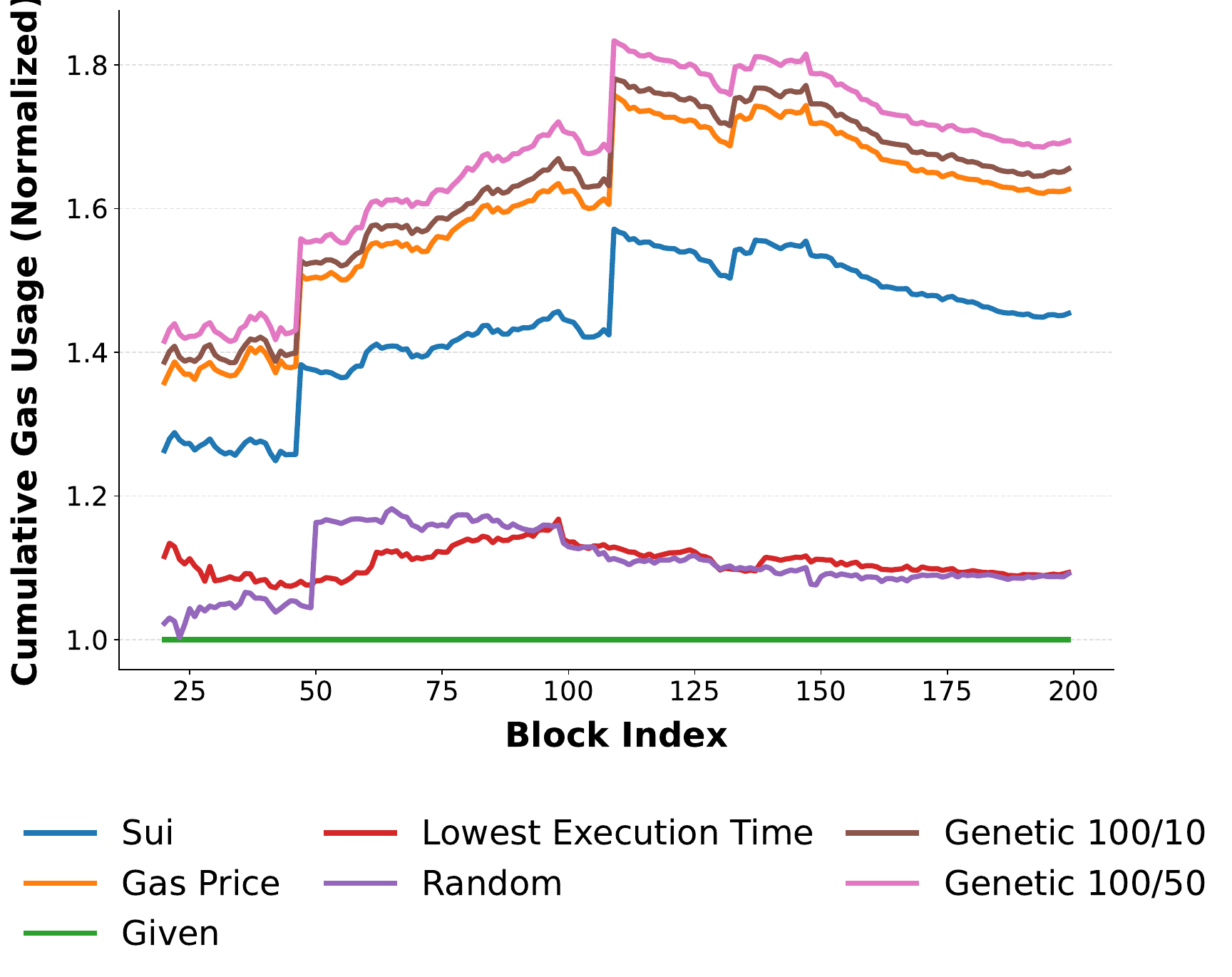}
        \caption{Ethereum dataset}
        \label{Validator profit under heavy load ETH}
    \end{subfigure}
    \caption{Total normalized validator profit under sustained congestion of 150 transactions per block for 200 blocks starting at block 20.}
\end{figure}

Figure~\ref{Validator profit under heavy load Sui} presents the results for the Sui dataset. All heuristic baselines except GP achieve similar performance. GP improves validator profit by about $20\%$ over the median heuristic, while the genetic sequencers provide a further $20\%$ gain. Increasing the number of generations from $10$ (GE10) to $50$ (GE50) yields only a modest additional improvement of roughly $5\%$, suggesting diminishing returns beyond $10$ generations.

For the Ethereum-based dataset, all of the parameters were identical, except for execution time, which was set to $1,250,000$ to provide the same $50-60$\% inclusion rate. The Ethereum dataset results, shown in Fig.~\ref{Validator profit under heavy load ETH}, display smaller improvements compared to the Sui dataset. GE10 had only a slight improvement over GP, while GE50  had the same proportional improvement, as it did for Sui. We believe that the main reason for this is the difference in gas price distribution. The gas prices for Sui have low variance as can be seen in Fig.~\ref{SUI data correlation}, while Ethereum's gas prices are much more varied as seen in Fig.~\ref{ETH data correlation}. This means that most of the profit is received by scheduling high-paying transactions. Since the first epochs of the genetic algorithm focus on global search with random initialization of transaction ordering, the algorithm is much less likely to find better sequences, because sequencing a high-paying transaction late will have a more severe negative impact on profit. This is also suggested by the Sui sequencer having a much better performance with the Ethereum-based than with the Sui dataset, since it prioritizes gas price. Due to elitism, the later epochs focus much more on local search, as the transaction set gets filled with similar transactions. 



\subsection{Transaction Spike}

\begin{figure}[t]
    \centering
    \begin{subfigure}[b]{0.45\textwidth}
        \centering
        \includegraphics[width=\textwidth]{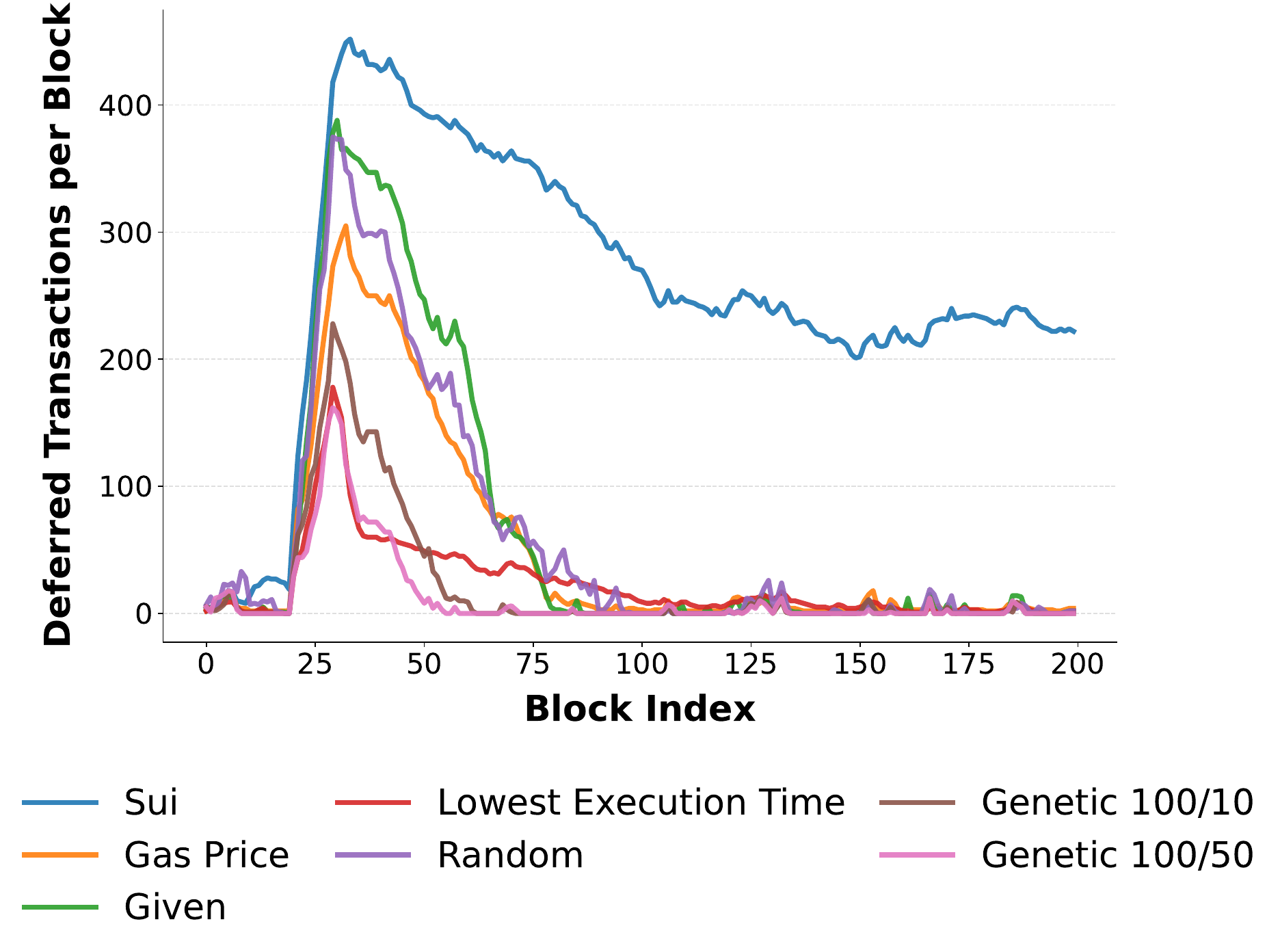}
        \caption{Sui dataset}
        \label{Number of deffered txs under a spike Sui}
    \end{subfigure}
    \hfill
    \begin{subfigure}[b]{0.45\textwidth}
        \centering
        \includegraphics[width=\textwidth]{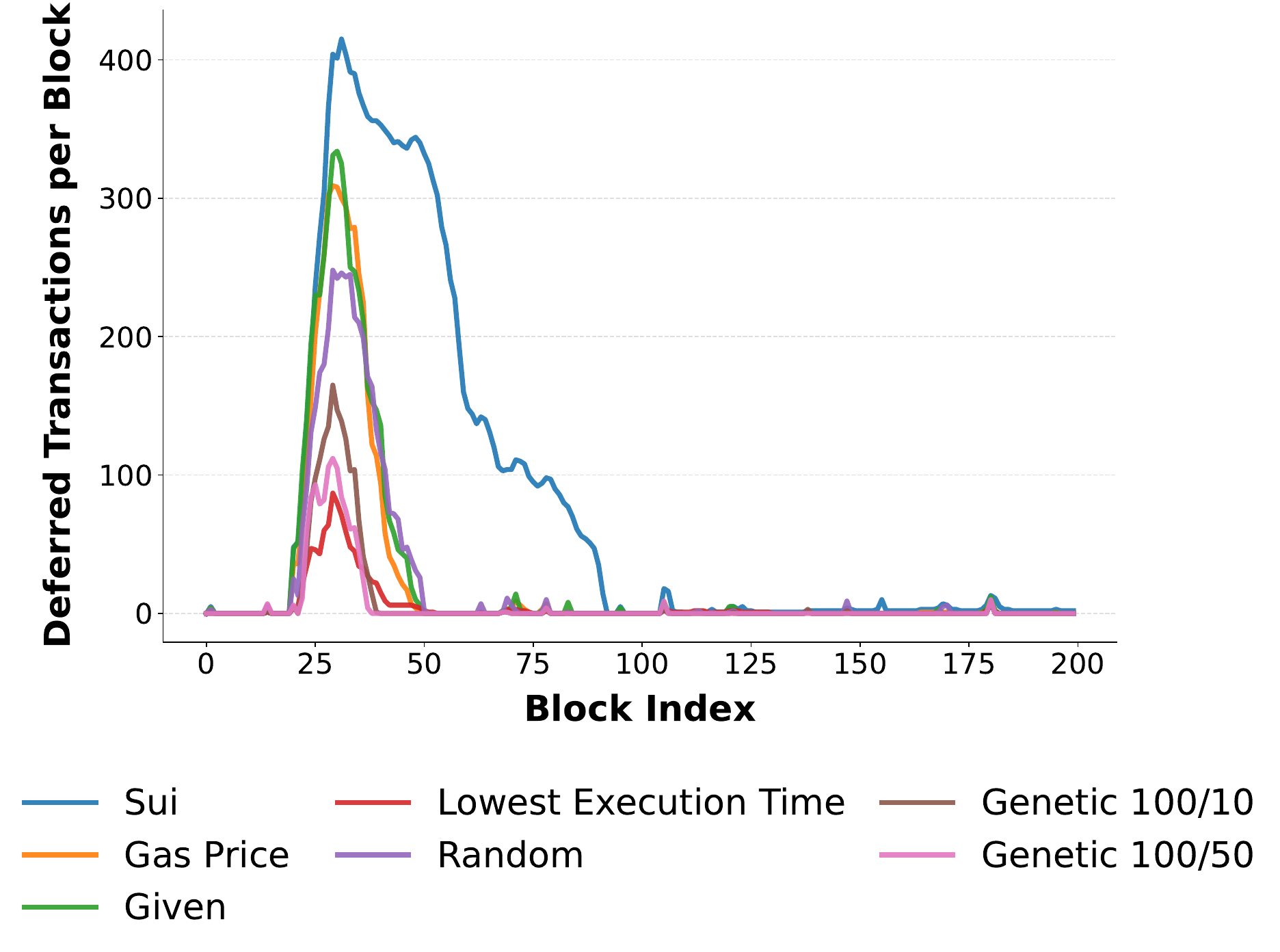}
        \caption{Ethereum dataset}
        \label{Number of deffered txs under a spike Ethereum}
    \end{subfigure}
    \caption{Total absolute number of transactions deferred under a spike in transaction volume from block $20$ to block $29$. Blocks $0-19$ contain $30$ transactions, blocks $20-29$ contain $100$ transactions, and blocks $30-199$ contain $30$ transactions.}
\end{figure}

To determine how well our genetic sequencing algorithm works in a scenario where congestion is not sustained but happens in sporadic bursts, we conducted an experiment where initially blocks contain a small number of transactions, then a high number of transactions for some time, and finally a small number of transactions again. We kept the number of executors at $4$ and the execution deadline at $62,500$ units for the Sui dataset. The maximum number of deferrals was set to $200$, so that no transactions would get canceled. The results showing the absolute number of deferred transactions for each block for Sui are shown in Fig.~\ref{Number of deffered txs under a spike Sui}. We see greater variations in the results of the heuristic baselines. The lack of gap filling prevented the Sui sequencer from recovering from the congestion. The Given and GP sequencers achieved similar results, recovering by block $130$. The LET and Random sequencers recovered around block $90$ with the LET sequencer showing a more rapid reduction of congestion, but then decreasing more slowly. Both of the genetic sequencers cleared congestion the fastest, with GE50 doing it by block $60$ and GE10 by block $70$.

For the Ethereum-based dataset the parameters were the same, with the only difference of execution deadline being, once again, set to $1,250,000$ units. The results are shown in Fig.~\ref{Number of deffered txs under a spike Ethereum}. For the Ethereum-based dataset the congestion was less severe and generally cleared much sooner than in the case of Sui. Besides that, the trend of how quickly sequencers manage to resolve congestion mirrors the Sui results with the exception of the Random sequencer performing on par with the Given and GP sequencers. This is likely because gas prices do not play a role in this scenario.



\subsection{Fair Ordering}

\begin{figure}[t]
    \centering
    \begin{subfigure}[b]{0.45\textwidth}
        \centering
        \includegraphics[width=\textwidth]{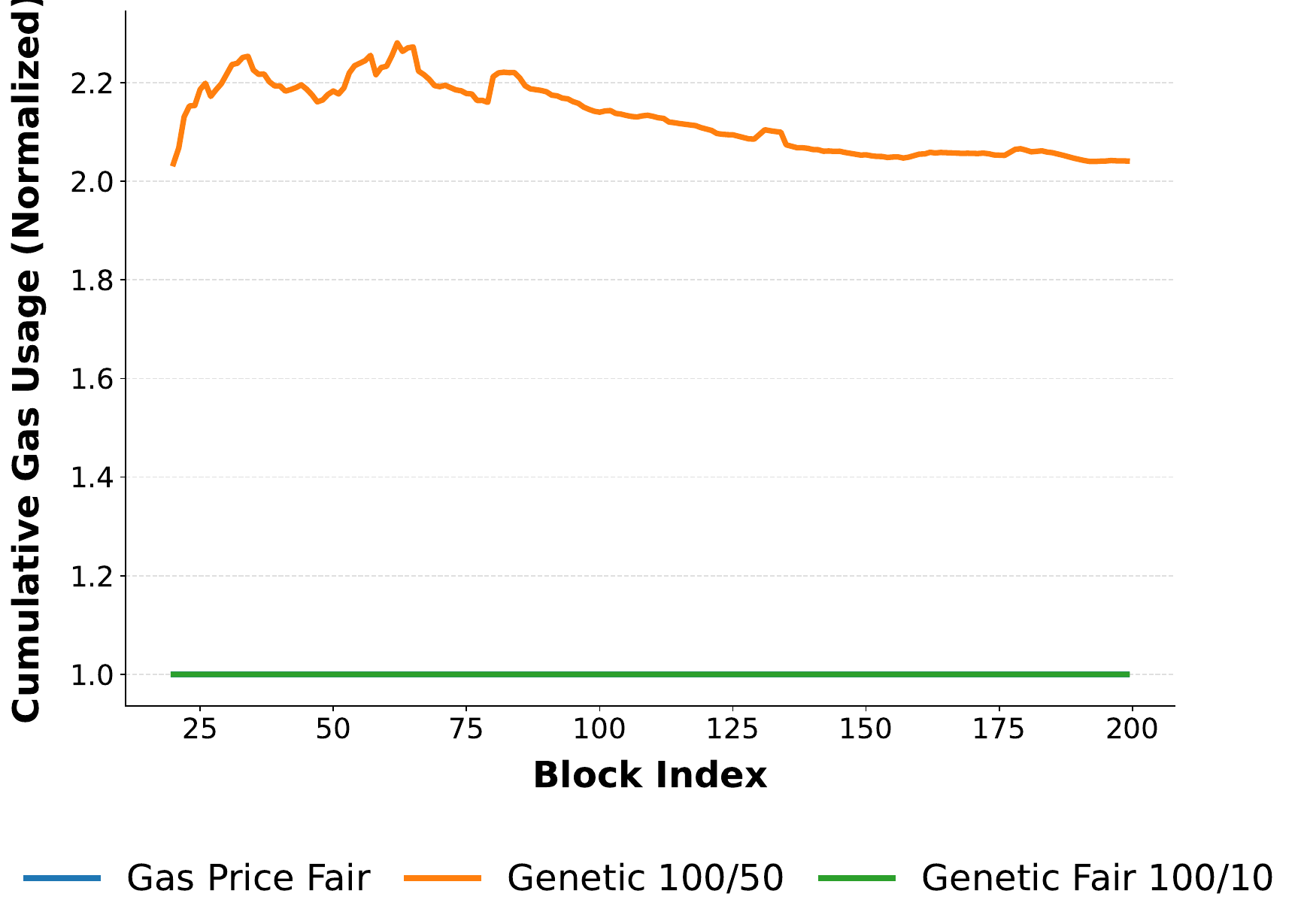}
        \caption{Sui dataset}
        \label{Validator profit under heavy load Sui fair}
    \end{subfigure}
    \hfill
    \begin{subfigure}[b]{0.45\textwidth}
        \centering
        \includegraphics[width=\textwidth]{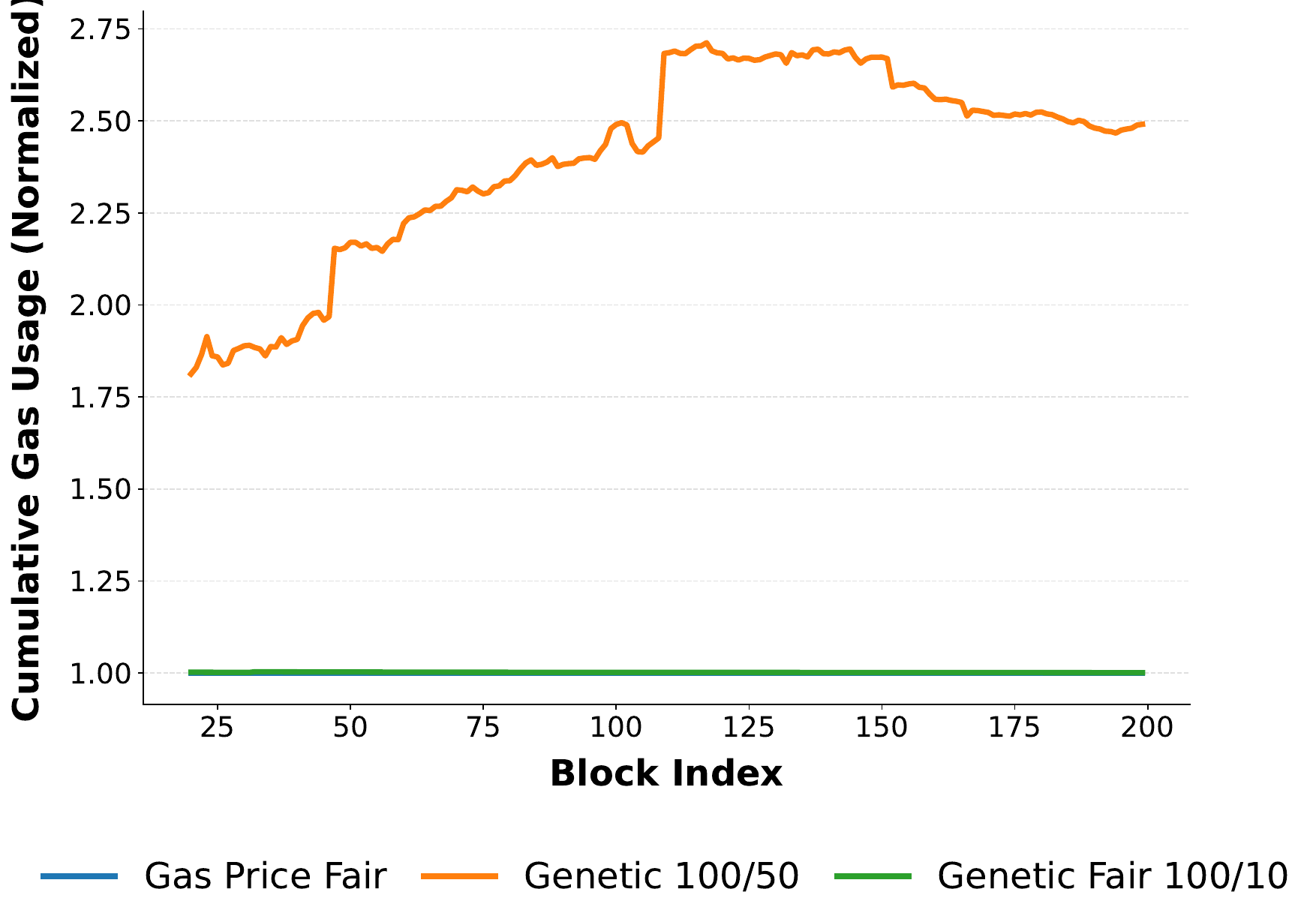}
        \caption{Ethereum dataset}
        \label{Validator profit under heavy load ETH fair}
    \end{subfigure}
    \caption{Total normalized validator profit under sustained congestion of $100$ transactions per block for $200$ blocks starting at block $20$. Sequencers include the fair-ordering schedulers and GE50, the most efficient non-fair scheduler.}
    \label{fig:validator-profit-heavy-load-fair}
\end{figure}

To evaluate the impact of fair ordering, we repeated the sustained-congestion experiment with fair-ordering constraints enabled. 
Figs.~\ref{Validator profit under heavy load Sui fair} and~\ref{Validator profit under heavy load ETH fair} respectively report the results for the Sui and for the Ethereum-based datasets. For both cases the fair variants of GP and GE10 produce the same results, suggesting that the fair ordering constraint does not leave sufficient room for reordering causally independent transactions, rendering the genetic algorithm ineffective. Additionally, this experiment highlights that under  sustained congestion, the addition of a fair ordering constraint reduces the profit by $50$ to $60$\%. We attribute this cost to the high rate of transaction interdependence in the dataset (see Fig.~\ref{SUI tx overlap}). 



    \subsection{Robustness to Execution Time Estimation Errors} \label{sec:robustness-to-execution-time}

The strongest assumption in our model is the availability of accurate execution-time estimates. To evaluate robustness to prediction errors, we repeated our experiments with perturbed execution times. Transactions were sequenced using the original execution times as predictions, but executed using perturbed values. Perturbations were generated by multiplying each execution time by a log-normal random variable: for each transaction, we sampled $X \sim \mathcal{N}(0,\,0.347^2)$ and used $e^X$ as the scaling factor, yielding a $2\sigma$ range of approximately $[0.5, 2]$. This configuration matches realistic prediction errors reported by Lima Cabral et al.~\cite{predictin_gas_price}, corresponding to an average absolute error of about $28\%$.

\begin{figure}[t]
    \centering
    \begin{subfigure}[b]{0.45\textwidth}
        \centering
        \includegraphics[width=\textwidth]{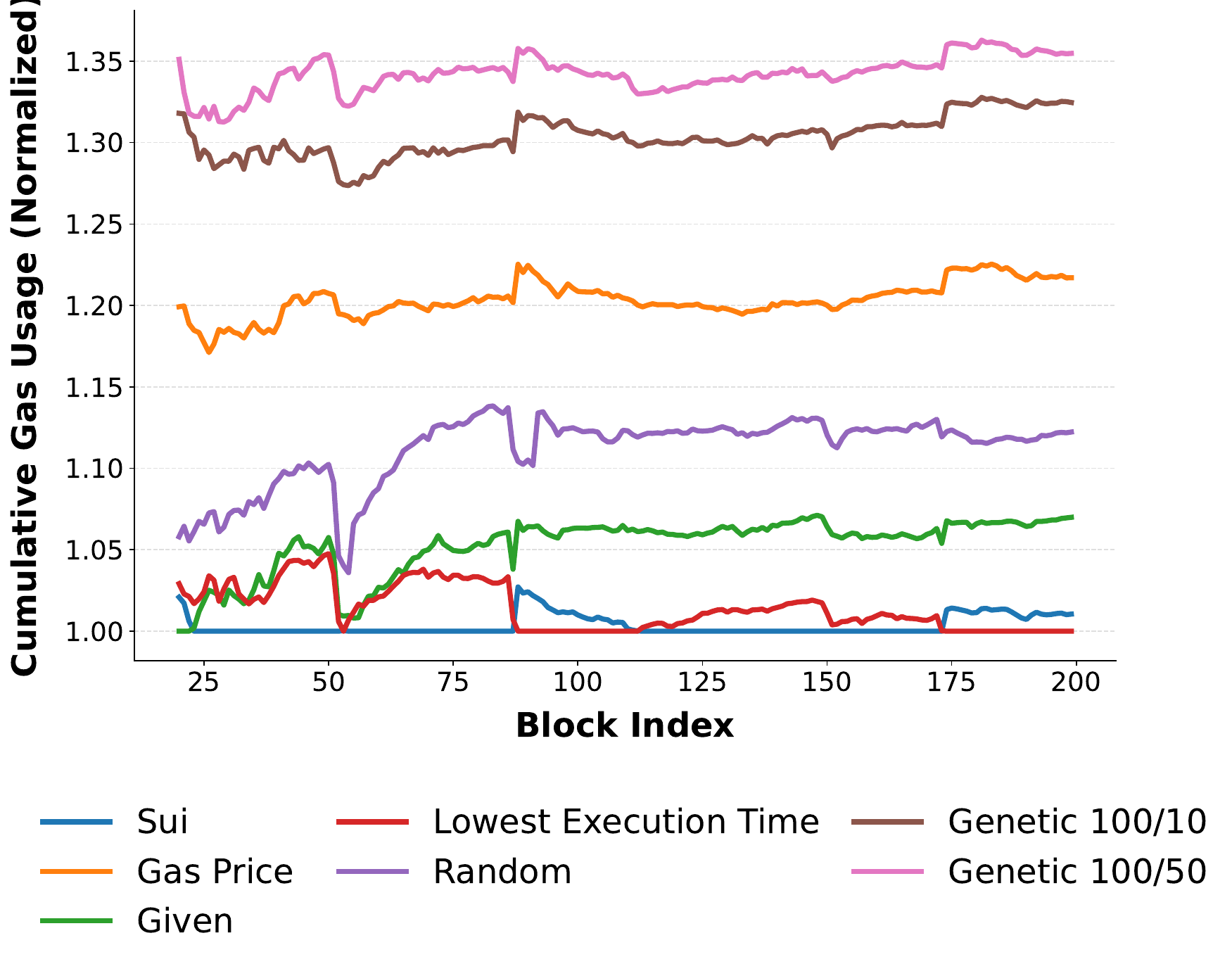}
        \caption{Sui dataset}
        \label{Validator profit under heavy load Sui preturbed}
    \end{subfigure}
    \hfill
    \begin{subfigure}[b]{0.45\textwidth}
        \centering
        \includegraphics[width=\textwidth]{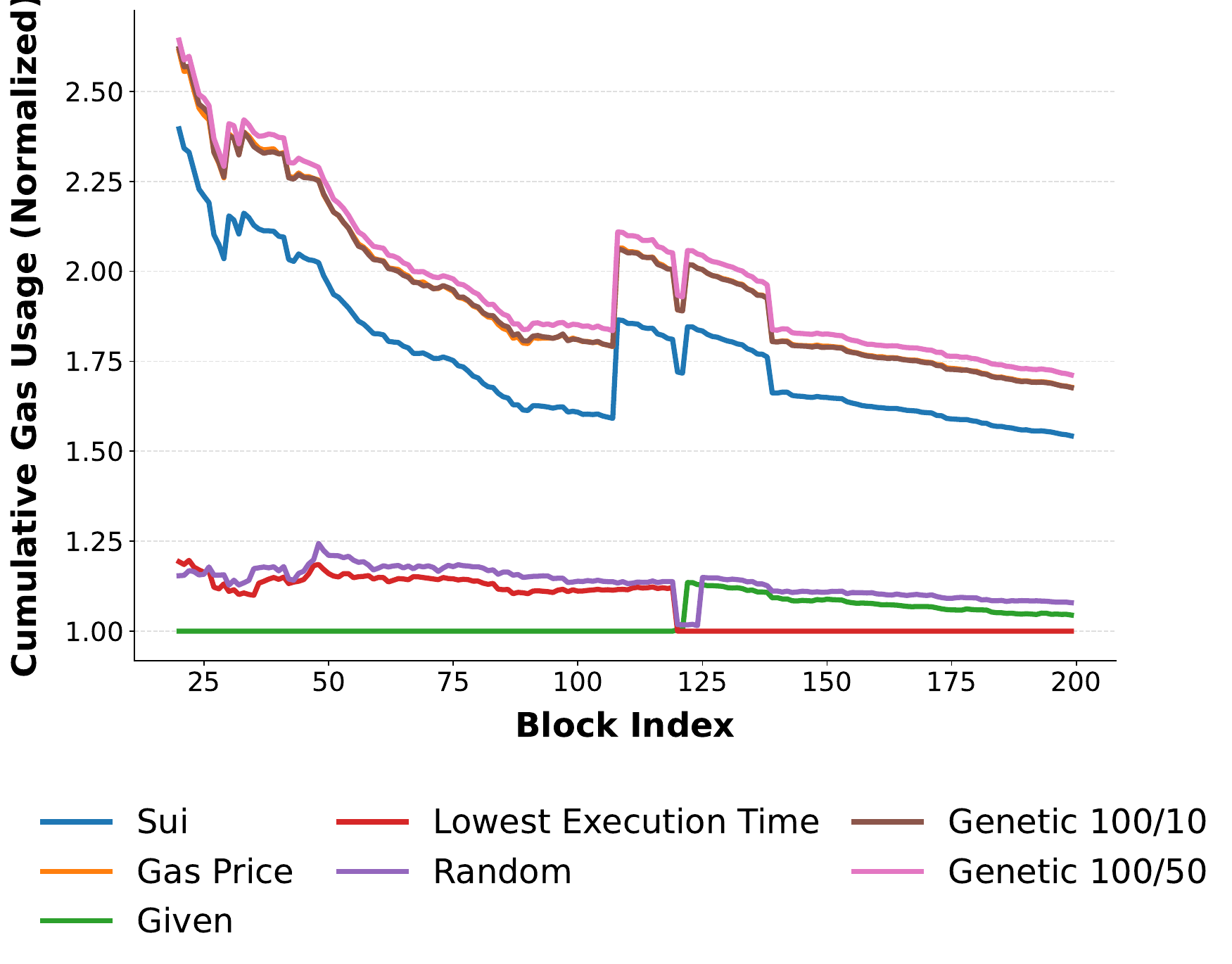}
        \caption{Ethereum dataset}
        \label{Validator profit under heavy load ETH preturbed}
    \end{subfigure}
    \caption{Total normalized validator profit under sustained congestion of $150$ transactions per block for $200$ blocks starting at block $20$ with perturbed transaction execution times.}
    \label{fig:validator-profit-heavy-load-perturbed}
\end{figure}

\begin{figure}[t]
    \centering
    \begin{subfigure}[b]{0.48\textwidth}
        \centering
        \includegraphics[width=\textwidth]{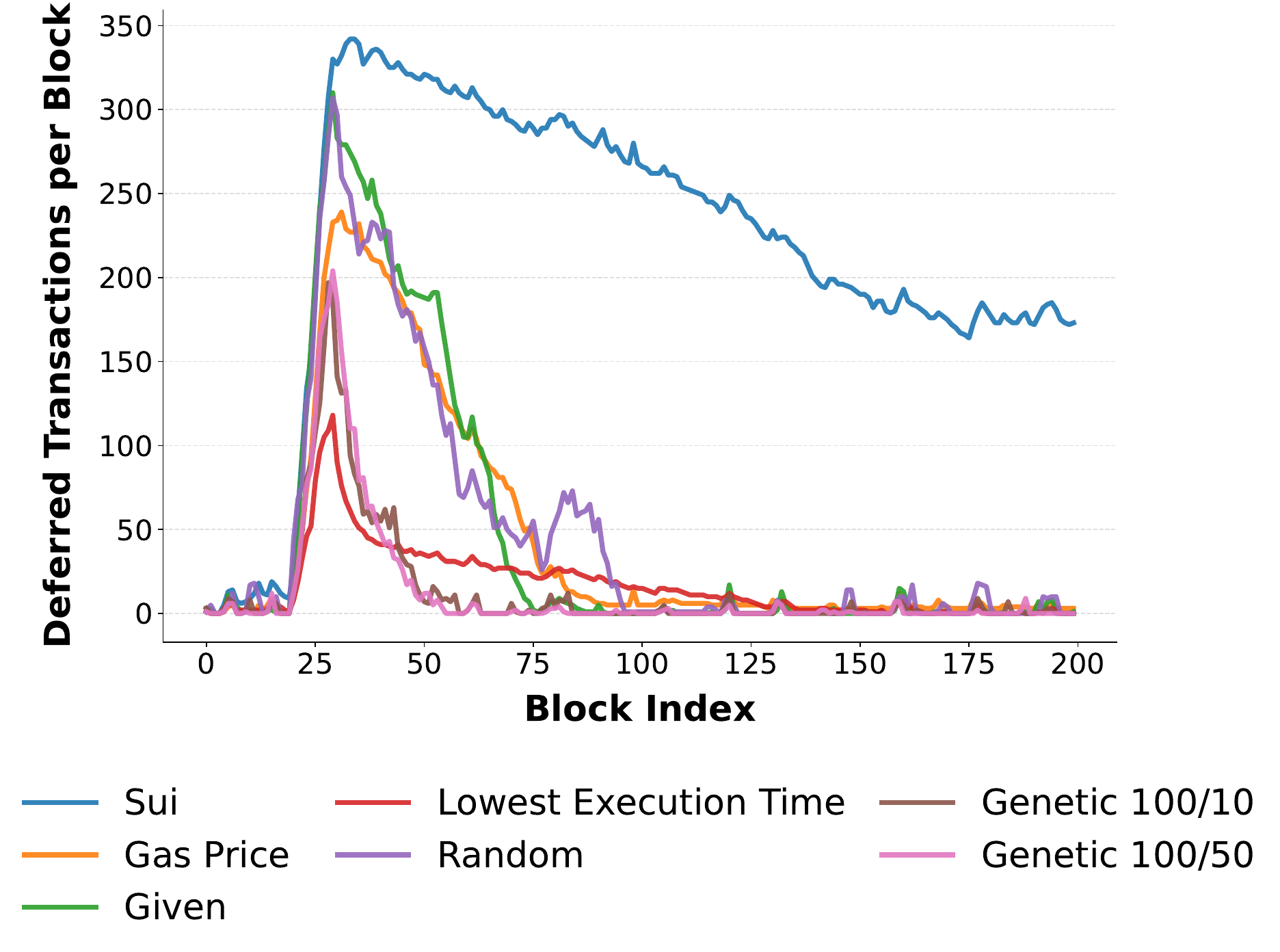}
        \caption{Sui dataset}
        \label{Number of deffered txs under a spike Sui preturbed}
    \end{subfigure}
    \hfill
    \begin{subfigure}[b]{0.48\textwidth}
        \centering
        \includegraphics[width=\textwidth]{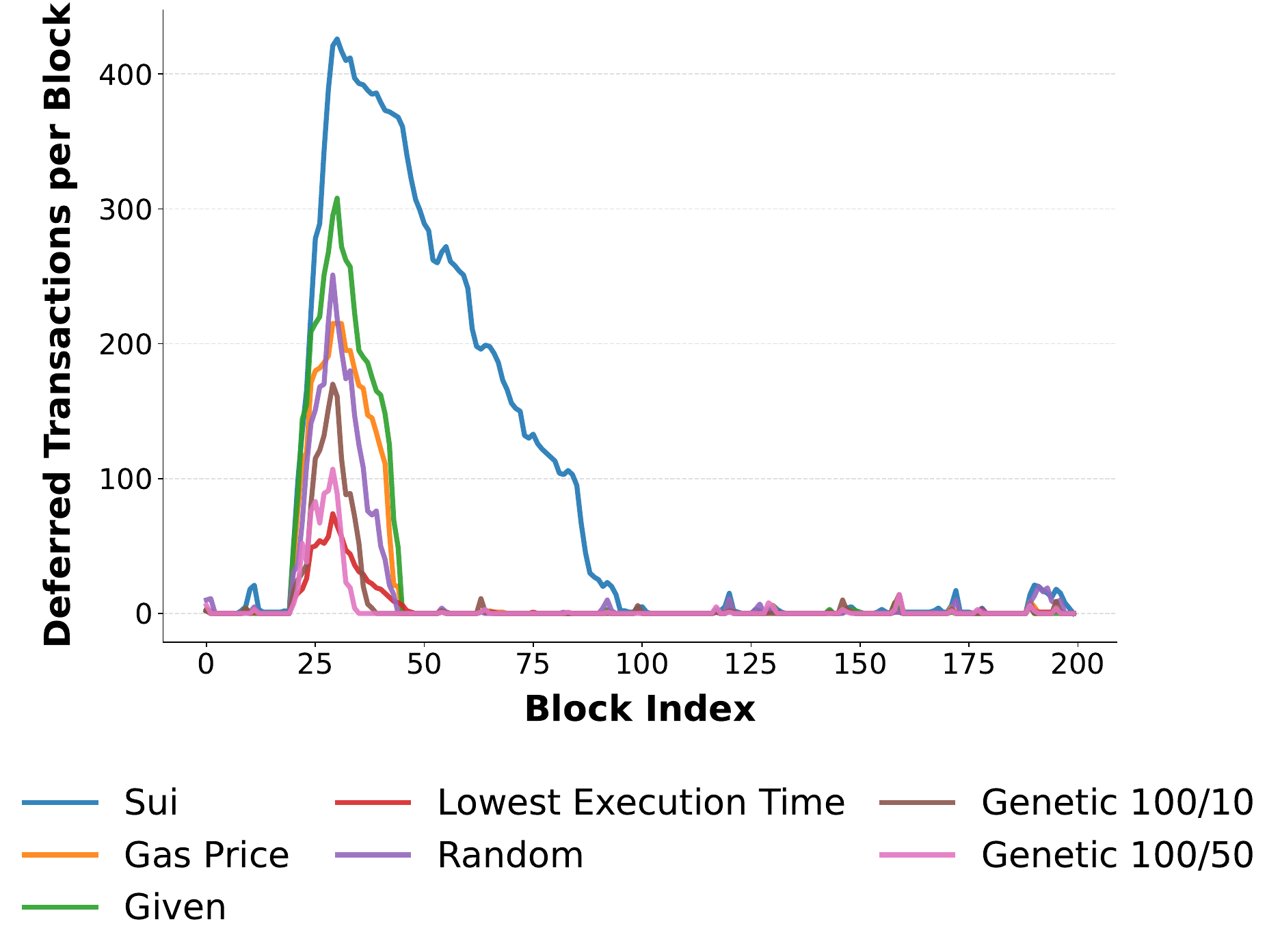}
        \caption{Ethereum dataset}
        \label{Number of deffered txs under a spike Ethereum preturbed}
    \end{subfigure}
    
    \caption{Total absolute number of transactions deferred under a spike (blocks $20-29$) for Sui and Ethereum datasets with perturbed transaction execution times. Baseline traffic is $30$ transactions/block, spiking to $150$ transactions/block during the spike.}
    \label{fig:deferred_txs_combined}
\end{figure}

%
Both the sustained congestion and the spike in congestion scenarios were tested with the addition of perturbation for both datasets. The sustained congestion for Sui seen in Fig.~\ref{Validator profit under heavy load Sui preturbed} and congestion spike for Sui and Ethereum seen in Figures~\ref{Number of deffered txs under a spike Sui preturbed} and~\ref{Number of deffered txs under a spike Ethereum preturbed} showed no statistically significant difference in speed of congestion clearance compared to the non-perturbed case. The sustained congestion for Ethereum seen in Fig.~\ref{Validator profit under heavy load ETH preturbed} shows that the genetic sequencers performed worse compared to their performance in the analogous experiment without perturbations, with only GE50 showing a very slight improvement over GP. 



\subsection{Throughput, Latency and Full Summary}

In this work, we optimize for validator profit, reflecting the transaction ordering objective used by most existing blockchains, rather than maximizing throughput or minimizing latency. Since we make no assumptions about the underlying hardware or blockchain implementation, we report normalized performance metrics. Specifically, throughput is defined as the average number of transactions executed within a time budget $d$, representing the maximum execution time on a single worker ($62{,}500$ units for the Sui dataset and $1{,}250{,}000$ units for the Ethereum-based dataset).


We measure latency as the time it takes from transaction sequencing until it has been executed in the units of $d$. Since our experiments simulated only the sequencing and execution steps, we assume no additional significant overhead at the dissemination and consensus steps and we do not account for the existing latency that they introduce. Additionally, since the simulation has no mechanisms to dynamically vary the time dedicated to sequencing, we assume that in each case sequencing runs until a worker becomes available.
Latency values below $1$ indicate early execution due to worker availability. Values between $1$ and $2$ indicate normal execution, and values above $2$ indicate that the transaction has been deferred. Canceled transactions are unaccounted for, meaning that the reported latency does not capture the full cost of congestion for affected users.

Results are given in \ifaft{Tables 3 and 4 in Appendix A of our extended paper~\cite{pugatsov2026blockchain}}\else{Tables~\ref{table:Table continued} and~\ref{table:Table spiked} (Appendix~\ref{app:additional_results})}\fi. The genetic sequencer consistently achieves the highest validator profit under sustained congestion, outperforming the GP baseline by respectively $16$\% and $4$\% for the Sui and Ethereum-based datasets. For the congestion spike scenario, the genetic sequencer clears congestion faster than the baseline, reducing the post-spike recovery time from $60$ to $25$ blocks on the Sui dataset and from $20$ to $8$ blocks on the ETH-based dataset, with consistent results even under perturbation. LET achieves the highest throughput at a significant cost to validator profit while the genetic sequencer consistently improves on the baseline throughput. Latency is broadly comparable across non-fair sequencers, though these figures should be interpreted cautiously, as the assumption of equal sequencing time is unrealistic. 
Fair ordering imposed a profit penalty of approximately $50$\% and failed to clear congestion on the Sui dataset.



\section{Related Work}
\label{sec:related_work}

We review four related literature areas. First, we examine transaction sequencing in production blockchains, ranging from full validator reordering to protocol-level congestion control. We then consider fair ordering approaches that favor transparency and manipulation resistance over congestion-aware optimization. Next, we review prior work framing sequencing as a discrete combinatorial optimization problem, with an overview provided in \ifaft{Table 5 in Appendix B of our full paper~\cite{pugatsov2026blockchain}}\else{Table~\ref{tab:sequencing_comparison} (Appendix~\ref{app:comparison_sequencing_approaches})}\fi. Finally, we briefly survey work on execution optimization.


\textbf{Sequencing in production blockchains.} 
In Bitcoin, transaction sequencing within a block was historically of limited importance. The main design goal was to maintain a consistent total order of blocks and ensure immutability unless an adversary controls a majority of hash power. Within a block, ordering was fully determined by the miner. Because Bitcoin uses a UTXO model, each transaction independently defines its inputs and outputs, making intra-block order irrelevant for validity. Only inclusion matters. In practice, transactions are ordered using ancestor fee rate policies, with dependency handling via Child-Pays-for-Parent (CPFP)~\cite{messias2021selfish}.

Ethereum initially followed a similar approach, but its account-based design makes ordering critical. Transactions read and modify shared state, so execution depends on ordering. This effect is strengthened by smart contracts~\cite{buterin2013ethereum}, which allow arbitrary state transitions. 
Early Ethereum clients such as Geth commonly ordered transactions greedily by gas price~\cite{kim2018measuring_geth_76_percent, ethereum_geth_worker_sort_greedy}. However, Daian et al.~\cite{daian2020flash} showed that validators can increase profit by reordering transactions and inserting their own, a phenomenon known as Miner Extractable Value (MEV). This led to MEV-boost and external block builders, with around 90\% adoption by 2022~\cite{oz2024wins}, increasing validator rewards by roughly 250–400\%~\cite{ethresear-mevboost-2024, mev_profit_boost_Mancino} but reducing transparency and decentralization. MEV is commonly extracted through private transaction submission or bundle-based ordering constraints~\cite{flashbots-mevshare-bundles}.

Sui is a DAG-based system that distinguishes between owned and shared objects, corresponding roughly to UTXO-like and account-like semantics. Owned-object transactions can be partially ordered without global sequencing, while shared objects require global ordering~\cite{SUI_OWNED_SHARED}. Originally, conflict resolution for owned objects relied on certification, but in current designs conflicting transactions must be resolved either during consensus or via sequencing. Sui’s sequencing layer also performs congestion control: transactions are first ordered by gas price and then selectively deferred based on estimated execution time and object-level capacity constraints~\cite{sui_congestion_control, sui_local_fee_markets}.

Solana has also modified its sequencing design in pursuit of high throughput~\cite{yakovenko2018solana}. Early designs used per-thread queues with local priority ordering, which produced partial and non-deterministic global order. Later versions introduced a global priority heap before thread assignment (v1.18). However, sequencing remains largely implementation defined, which has enabled MEV-oriented validator clients such as Jito~\cite{helius-solana-local-fee-markets}.
This illustrates a recurring pattern: when sequencing is not specified at the protocol level, economic incentives tend to drive systems toward opaque, profit-maximizing ordering strategies.

\textbf{Fair ordering.} 
The lack of transparency in transaction ordering and the rise of MEV-boosting have motivated research on fair transaction ordering~\cite{kelkar2022order}. Its goal is to align transaction order as closely as possible with real submission times. Two main approaches exist. Ordering linearizability, introduced in Pomp\=e~\cite{zhang2020byzantinepompetimestamp}, uses validator-provided timestamps and orders transactions by the median of a subset of its timestamps. Batch order fairness, as in Themis~\cite{kelkar2023themisbatch}, aggregates full per-validator orderings into a global sequence.
Fair ordering improves protocol-level fairness and reduces MEV opportunities, but it imposes a rigid global order, lacks congestion control, and ignores validator incentives.

\textbf{Combinatorial optimization for sequencing.}
Under deterministic execution, sequencing reduces to block construction. With sequential execution and position-independent execution times, the problem becomes a knapsack instance and is therefore NP-hard~\cite{sequencing_as_knapsack}. Karmegam et al.~\cite{karmegam2026exploitingmulticoreparallelismblockchain} formulate block construction as a mixed-integer program maximizing validator revenue under predefined dependencies and implicit fair-ordering constraints. Their heuristic performs similarly to a gas-price greedy baseline, consistent with our finding that genetic optimization offers little benefit when fairness constraints are enforced. Conthereum~\cite{chahoki2025conthereum} models sequencing as a job-shop scheduling problem, optimizing makespan under unrestricted reordering while requiring all transactions to fit within a block. In contrast, we consider continuous execution with overlapping blocks, evaluate robustness to execution-time prediction errors, and study both unconstrained and fair transaction ordering.

\textbf{Execution.} 
Most prior work targets improvements in the execution layer. Pilotfish~\cite{pilotfish} is a distributed engine that executes transactions in order while dynamically assigning independent ones to workers, enabling parallelism under a crash-recovery model with low communication overhead.
Extensions of Hyperledger Fabric, such as HTFabric and Fabric++~\cite{HTFabric, Fabric_plus_plus}, further increase parallelism by reordering transactions during execution. While this improves throughput, it biases against interdependent transactions.
Other approaches avoid requiring predeclared object sets. Anjana et al.~\cite{anjana2019efficientframeworkoptimisticconcurrent} use software transactional memory (STM) to speculatively execute transactions in parallel with rollback on conflicts. Block-STM~\cite{Block-STM} refines dependency estimates after rollbacks, reducing redundant execution.

\section{Discussion}

\textbf{Gas price.}  
Users do not seem to price the congestion they impose in their transactions, as shown by the weak correlation between gas price and the number of accessed objects (Figs.~\ref{SUI data correlation} and~\ref{ETH data correlation}). Furthermore, we assume a static, user-specified gas price throughout execution. Incorporating dynamic, multi-dimensional pricing mechanisms tailored to parallel execution environments~\cite{acilan2025transaction,cryptoeprint:2026/649} is a promising direction for future work.

\textbf{Fair ordering constraints and conflict resolution.} 
We derived causal order from total ordering. The observed 50\% to 60\% profit penalty under sustained congestion highlights the strictness of enforcing this causal chain, at least with the algorithms and datasets we considered. Whether this penalty is verified with all schedulers and datasets is an open question. 
Alternatively, batch order fairness~\cite{kelkar2023themisbatch,putnik2026herring} introduces dependency cycles that could offer scheduling flexibility, though such cycles are rare in practice. Exploring alternative optimization methods like GRASP or simulated annealing remains future work.



\textbf{External gameability and input resilience. } 
Adversaries cannot game placement via gas bidding without incurring direct execution fees.
Furthermore, our evaluation in Section~\ref{sec:robustness-to-execution-time} demonstrates that our genetic algorithm remains robust against significant execution-time estimation errors. 
To prevent the case where adversaries declare unused shared objects to trigger inaccurate deferrals, our framework seeds optimization with gas price heuristics to protect higher-paying honest transactions.
A possible direction might be the development of native localized fee markets targeting specific shared objects, introducing dynamic fee penalties for declared-but-unaccessed objects or localized pricing mechanisms based on state-declaration accuracy~\cite{acilan2025transaction, cryptoeprint:2026/649}.

\textbf{Real-Time Sequencing.} While prior work~\cite{chahoki2025conthereum} found genetic algorithms and constraint solvers too slow for online sequencing, our approach uses an anytime algorithm with strict execution deadlines. 
Seeding the initial population with a gas-price heuristic provides an immediately deployable baseline schedule, while hard execution deadlines  interrupt search to ensure continuous worker utilization.

\textbf{Architectural assumptions.} 
Our model assumes deterministic sequencing over committed consensus batches. While single-leader block builders could act as privileged clients to censor transactions,  our target deployment environment relies on DAG-based consensus (\cite{danezis2022narwhal, spiegelman2022bullshark, babel2025mysticeti}). Because transactions are broadcast to multiple nodes and committed to consensus blocks before sequencing, individual validators cannot selectively exclude transactions during optimization without violating state consistency.

\section{Conclusion}
\label{chapter:conclusion}

We have formalized a blockchain-independent model for evaluating the impact of transaction ordering on validator profit and congestion clearance speed. We extended this model to account for continuity of execution and fair ordering.
We then proposed a genetic ordering algorithm that takes into account transaction execution time estimates and object conflict information.
This approach outperforms the heuristic baseline, increasing validator profit by approximately $15$\%, and improves congestion control in a high congestion scenario. We have also shown that the performance of this approach does not degrade significantly under realistic execution time estimation errors for the purpose of congestion control. Performance also does not degrade significantly in terms of profit maximization when gas prices exhibit little variation.
We finally integrated a fair ordering constraint and found that under such an assumption, sequencing provides no benefit. Introducing fair ordering constraints lead validators to lose around $50$ to $60$\% of their profit under sustained congestion.



\bibliography{report.bib}

\ifaft{}
\else
\appendix

\section{Additional Experimental Results}
\label{app:additional_results}

Tables~\ref{table:Table continued} and~\ref{table:Table spiked} summarize the performance of all evaluated sequencing algorithms under two representative congestion scenarios. Table~\ref{table:Table continued} reports results for sustained congestion, where every block contains more transactions than can be executed, while Table~\ref{table:Table spiked} considers a temporary surge in demand followed by a return to normal load.

For each experiment, we report throughput, latency, and a scenario-specific metric that captures the primary optimization objective. Under sustained congestion, this metric is the normalized amount of gas used, which reflects the relative validator revenue. Under transient congestion, we instead report the number of blocks required to fully recover, defined as the number of consecutive blocks after the spike until no transactions remain deferred. Results are shown for both the Sui and Ethereum-based datasets, with and without perturbations in transaction execution times.

\begin{table*}
\centering
\caption{Results of experiments with continued congestion. Throughput is measured as the number of transactions that are scheduled in a block. Validator profit is normalized to the worst-performing sequencer, and is the metric we aim to maximize. Latency is measured as a multiple of the number of blocks from the start of sequencing to the end of execution.}
\label{table:Table continued}
\resizebox{\columnwidth}{!}{%
\begin{tabular}{l|c|c|c}
\hline
Sequencer & Throughput $\uparrow$ & Profit (norm.) $\uparrow$ & Latency $\downarrow$ \\
\hline
\multicolumn{4}{c}{\textbf{Sui dataset}} \\ \hline
\multicolumn{4}{l}{\textit{No Perturbation}} \\ \hline
Sui & 70.32 $\pm$ 1.18 & 1.41 $\pm$ 0.03 & 1.27 $\pm$ 0.02 \\
Gas Price  & 86.60 $\pm$ 1.31 & 1.76 $\pm$ 0.03 & 1.20 $\pm$ 0.02 \\
Given & 76.58 $\pm$ 1.23 & 1.49 $\pm$ 0.02 & 1.02 $\pm$ 0.02 \\
Lowest Execution Time & \textbf{101.58 $\pm$ 0.80} & 1.49 $\pm$ 0.01 & \textbf{0.90 $\pm$ 0.01} \\
Random & 77.29 $\pm$ 1.32 & 1.53 $\pm$ 0.02 & 2.04 $\pm$ 0.02 \\
Fair Highest gas price & 52.95 $\pm$ 0.96 & 1.00 $\pm$ 0.02 & 2.76 $\pm$ 0.10 \\
Genetic 100/10 (this work) & 95.59 $\pm$ 1.16 & 1.97 $\pm$ 0.03 & 1.56 $\pm$ 0.02 \\
Genetic 100/50 (this work) & 99.89 $\pm$ 1.12 & \textbf{2.04 $\pm$ 0.03} & 1.45 $\pm$ 0.02 \\
Fair Genetic 100/10 (this work) & 52.96 $\pm$ 0.95 & 1.00 $\pm$ 0.02 & 2.60 $\pm$ 0.11 \\
\hline
\multicolumn{4}{l}{\textit{With Perturbations}} \\ \hline
Sui & 74.51 $\pm$ 1.10 & 1.42 $\pm$ 0.04 & 1.10 $\pm$ 0.02 \\
Gas Price  & 88.80 $\pm$ 1.20 & 1.71 $\pm$ 0.04 & 1.10 $\pm$ 0.02 \\
Given & 81.69 $\pm$ 1.13 & 1.50 $\pm$ 0.03 & \textbf{0.93 $\pm$ 0.02} \\
Lowest Execution Time & \textbf{109.81 $\pm$ 0.91} & 1.40 $\pm$ 0.02 & 1.10 $\pm$ 0.01 \\
Random & 83.53 $\pm$ 1.24 & 1.58 $\pm$ 0.03 & 1.83 $\pm$ 0.02 \\
Fair Highest gas price & 59.53 $\pm$ 0.97 & 1.00 $\pm$ 0.01 & 2.03 $\pm$ 0.07 \\
Genetic 100/10 (this work) & 93.36 $\pm$ 1.17 & 1.86 $\pm$ 0.04 & 1.53 $\pm$ 0.02 \\
Genetic 100/50 (this work) & 96.69 $\pm$ 1.23 & \textbf{1.90 $\pm$ 0.04} & 1.54 $\pm$ 0.02 \\
Fair Genetic 100/10 (this work) & 59.54 $\pm$ 0.97 & 1.00 $\pm$ 0.01 & 2.03 $\pm$ 0.07 \\
\hline
\multicolumn{4}{c}{\textbf{ETH dataset}} \\ \hline
\multicolumn{4}{l}{\textit{No Perturbation}} \\ \hline
Sui & 68.99 $\pm$ 1.22 & 2.14 $\pm$ 0.15 & 1.35 $\pm$ 0.02 \\
Gas Price  & 90.08 $\pm$ 1.52 & 2.39 $\pm$ 0.16 & 1.17 $\pm$ 0.02 \\
Given & 90.89 $\pm$ 1.90 & 1.47 $\pm$ 0.07 & 1.07 $\pm$ 0.01 \\
Lowest Execution Time & \textbf{125.83 $\pm$ 1.07} & 1.61 $\pm$ 0.07 & \textbf{1.05 $\pm$ 0.01} \\
Random & 89.02 $\pm$ 1.66 & 1.61 $\pm$ 0.07 & 2.17 $\pm$ 0.02 \\
Fair Highest gas price & 62.67 $\pm$ 1.57 & 1.00 $\pm$ 0.05 & 3.09 $\pm$ 0.09 \\
Genetic 100/10 (this work) & 93.50 $\pm$ 1.45 & 2.43 $\pm$ 0.15 & 1.26 $\pm$ 0.02 \\
Genetic 100/50 (this work) & 97.47 $\pm$ 1.55 & \textbf{2.49 $\pm$ 0.16} & 1.23 $\pm$ 0.02 \\
Fair Genetic 100/10 (this work) & 62.75 $\pm$ 1.57 & 1.00 $\pm$ 0.05 & 2.99 $\pm$ 0.09 \\
\hline
\multicolumn{4}{l}{\textit{With Perturbations}} \\ \hline
Sui & 74.67 $\pm$ 1.22 & 2.01 $\pm$ 0.27 & 1.28 $\pm$ 0.02 \\
Gas Price  & 96.60 $\pm$ 1.55 & 2.19 $\pm$ 0.27 & 1.10 $\pm$ 0.02 \\
Given & 96.66 $\pm$ 1.69 & 1.36 $\pm$ 0.18 & 1.08 $\pm$ 0.02 \\
Lowest Execution Time & \textbf{129.37 $\pm$ 0.90} & 1.31 $\pm$ 0.13 & \textbf{1.03 $\pm$ 0.01} \\
Random & 96.56 $\pm$ 1.53 & 1.41 $\pm$ 0.16 & 2.03 $\pm$ 0.02 \\
Fair Highest gas price & 68.92 $\pm$ 1.45 & 1.00 $\pm$ 0.16 & 2.79 $\pm$ 0.08 \\
Genetic 100/10 (this work) & 96.95 $\pm$ 1.45 & 2.19 $\pm$ 0.27 & 1.27 $\pm$ 0.02 \\
Genetic 100/50 (this work) & 100.05 $\pm$ 1.49 & \textbf{2.23 $\pm$ 0.27} & 1.28 $\pm$ 0.02 \\
Fair Genetic 100/10 (this work) & 68.95 $\pm$ 1.45 & 1.00 $\pm$ 0.16 & 2.78 $\pm$ 0.07 \\
\hline
\end{tabular}
}
\end{table*}

\begin{table*}
\centering

\caption{Results of experiments with a spike in the number of transactions. Throughput is measured as the number of transactions that are scheduled in a block. Number of blocks to recover from congestion represents the count of consecutive blocks after congestion ends until a block is produced with no deferred transactions. Latency is measured as a multiple of the number of blocks from the start of sequencing to the end of execution.}
\label{table:Table spiked}

\resizebox{\columnwidth}{!}{%
\begin{tabular}{l|c|c|c}

\hline
Sequencer & Throughput $\uparrow$ & \# Blocks to recover from congestion $\downarrow$ & Latency $\downarrow$ \\
\hline
\multicolumn{4}{c}{\textbf{Sui dataset}} \\ \hline
\multicolumn{4}{l}{\textit{No Perturbation}} \\ \hline
Sui & 32.39 $\pm$ 0.57 & Did not recover & 3.82 $\pm$ 0.30 \\
Gas Price  & 33.48 $\pm$ 0.74 & 60 & 2.20 $\pm$ 0.19 \\
Given & 33.49 $\pm$ 0.69 & 57 & 2.38 $\pm$ 0.27 \\
Lowest Execution Time & 33.49 $\pm$ 0.84 & 162 & 1.34 $\pm$ 0.08 \\
Random & 33.49 $\pm$ 0.76 & 66 & 2.52 $\pm$ 0.18 \\
Fair Highest gas price & 27.95 $\pm$ 0.74 & Did not recover & 21.68 $\pm$ 0.84 \\
Genetic 100/10 (this work) & \textbf{33.50 $\pm$ 0.80} & 32 & 1.40 $\pm$ 0.09 \\
Genetic 100/50 (this work) & \textbf{33.50 $\pm$ 0.88} & \textbf{25} & \textbf{1.19 $\pm$ 0.07} \\
Fair Genetic 100/10 (this work) & 27.95 $\pm$ 0.74 & Did not recover & 21.26 $\pm$ 0.86 \\
\hline
\multicolumn{4}{l}{\textit{With Perturbations}} \\ \hline
Sui & 32.63 $\pm$ 0.65 & Did not recover & 3.94 $\pm$ 0.27 \\
Gas Price  & 33.48 $\pm$ 0.76 & 59 & 2.13 $\pm$ 0.18 \\
Given & 33.49 $\pm$ 0.70 & 62 & 2.09 $\pm$ 0.21 \\
Lowest Execution Time & \textbf{33.50 $\pm$ 0.91} & 126 & \textbf{1.23 $\pm$ 0.06} \\
Random & \textbf{33.50 $\pm$ 0.81} & 88 & 2.32 $\pm$ 0.15 \\
Fair Highest gas price & 26.64 $\pm$ 0.78 & Did not recover & 28.16 $\pm$ 1.17 \\
Genetic 100/10 (this work) & \textbf{33.50 $\pm$ 0.91} & 28 & 1.29 $\pm$ 0.06 \\
Genetic 100/50 (this work) & \textbf{33.50 $\pm$ 0.84} & \textbf{25} & 1.24 $\pm$ 0.06 \\
Fair Genetic 100/10 (this work) & 26.64 $\pm$ 0.78 & Did not recover & 28.08 $\pm$ 1.18 \\
\hline
\multicolumn{4}{c}{\textbf{ETH dataset}} \\ \hline
\multicolumn{4}{l}{\textit{No Perturbation}} \\ \hline
Sui & 33.49 $\pm$ 0.64 & 62 & 2.80 $\pm$ 0.31 \\
Gas Price  & \textbf{33.50 $\pm$ 0.80} & 20 & 1.28 $\pm$ 0.11 \\
Given & \textbf{33.50 $\pm$ 0.79} & 20 & 1.35 $\pm$ 0.12 \\
Lowest Execution Time & \textbf{33.50 $\pm$ 0.96} & 23 & \textbf{0.86 $\pm$ 0.04} \\
Random & \textbf{33.50 $\pm$ 0.83} & 21 & 1.35 $\pm$ 0.10 \\
Fair Highest gas price & \textbf{33.50 $\pm$ 0.84} & 81 & 3.64 $\pm$ 0.33 \\
Genetic 100/10 (this work) & \textbf{33.50 $\pm$ 0.89} & 10 & 1.00 $\pm$ 0.06 \\
Genetic 100/50 (this work) & \textbf{33.50 $\pm$ 0.97} & \textbf{8} & 0.98 $\pm$ 0.04 \\
Fair Genetic 100/10 (this work) & \textbf{33.50 $\pm$ 0.84} & 81 & 3.49 $\pm$ 0.33 \\
\hline
\multicolumn{4}{l}{\textit{With Perturbations}} \\ \hline
Sui & \textbf{33.50 $\pm$ 0.62} & 72 & 3.14 $\pm$ 0.34 \\
Gas Price  & \textbf{33.50 $\pm$ 0.88} & 15 & 1.19 $\pm$ 0.11 \\
Given & \textbf{33.50 $\pm$ 0.84} & 15 & 1.27 $\pm$ 0.13 \\
Lowest Execution Time & \textbf{33.50 $\pm$ 0.97} & 18 & \textbf{0.89 $\pm$ 0.03} \\
Random & \textbf{33.50 $\pm$ 0.83} & 14 & 1.21 $\pm$ 0.07 \\
Fair Highest gas price & 33.39 $\pm$ 0.96 & 82 & 4.96 $\pm$ 0.38 \\
Genetic 100/10 (this work) & \textbf{33.50 $\pm$ 0.91} & 9 & 1.01 $\pm$ 0.05 \\
Genetic 100/50 (this work) & \textbf{33.50 $\pm$ 0.99} & \textbf{5} & 0.95 $\pm$ 0.03 \\
Fair Genetic 100/10 (this work) & 33.39 $\pm$ 0.96 & 82 & 4.98 $\pm$ 0.38 \\
\hline
\end{tabular}
}
\end{table*}

\section{Comparison of Sequencing Approaches}
\label{app:comparison_sequencing_approaches}

Table~\ref{tab:sequencing_comparison} compares transaction sequencing approaches, and gives an overview of the criteria used for transaction ordering and evaluates to which extent the resulting orderings are deterministic, are resistant to manipulation, specifically preventing validators from arbitrarily reordering, including, or excluding transactions to their own advantage, and are aligned with validator incentives.


\begin{table}[htbp]
\centering
\caption{Comparison of different transaction sequencing approaches. We refer the reader to Sec.~\ref{sec:related_work} for a detailed comparison between Conthereum and our work (key differences include sequencing continuity, fair ordering support, and noisy execution times).}
\label{tab:sequencing_comparison}
\footnotesize
\renewcommand{\arraystretch}{1.4}
\setlength{\tabcolsep}{3pt}

\rowcolors{2}{gray!10}{white}

\begin{tabularx}{\columnwidth}{
    >{\raggedright\arraybackslash}p{2.4cm}
    >{\raggedright\arraybackslash}X
    c
    c
    c
}
\rowcolor{white}
 &
\textbf{Ordering Criteria} &
\makecell{\textbf{Determinism}} &
\makecell{\textbf{Manipulation}\\\textbf{Resistance}} &
\makecell{\textbf{Validator}\\\textbf{Alignment}} \\  
\midrule
Bitcoin &
Child-Pays-For-Parent (CPFP) &
High &
Moderate &
High \\

Eth (pre-MEV) &
Gas price &
High &
Low &
Moderate \\

Eth (MEV-Boost) &
Builder bid &
Low &
Low &
Very High \\

Sui &
Gas price &
High &
High &
Moderate \\

Solana ($<$v1.18) &
Priority fee (per-thread) &
Low &
Low &
Low \\

Solana ($>$v1.18) &
Priority fee (global heap) or builder bid &
Moderate &
Moderate &
High \\

Fair Ordering &
Arrival timestamps &
Very High &
Very High &
Low \\

This work and Conthereum~\cite{chahoki2025conthereum} &
Gas price, execution time and touched objects &
High &
High &
High \\
\bottomrule
\end{tabularx}

\rowcolors{2}{}{}
\end{table}

\fi

\end{document}